\documentclass[journal]{IEEEtran}
\usepackage{graphics,graphicx}
\usepackage{latexsym}
\usepackage{algorithm,algorithmic}
\usepackage{array}
\usepackage{mdwmath}
\makeatletter
\newcommand*{\rom}[1]{\expandafter\@slowromancap\romannumeral #1@}
\usepackage{mdwtab}
\usepackage{eqparbox}
\usepackage{stfloats}
\usepackage{latexsym}
\usepackage{amssymb}
\usepackage{amsmath}
\usepackage{graphicx}
\usepackage{csquotes}
\usepackage{balance}
\usepackage[justification=centering]{caption}
\DeclareGraphicsExtensions{.eps,.pdf,.gif,.png,.jpg}
\usepackage{color}
\usepackage[breaklinks=true]{hyperref}
\usepackage{breakcites}
\usepackage{url}  
\usepackage{amsthm}
\usepackage{multirow}
\usepackage{color}
\usepackage{epstopdf}
\usepackage{endnotes}
\usepackage{framed}
\usepackage{cool} 
\usepackage{enumerate} 
\usepackage{url}
\usepackage[subnum]{cases}
\usepackage[nocomma]{optidef}
\usepackage{etoolbox}
\usepackage{cite}

\ifCLASSINFOpdf
\else
\fi
\begin{document}
%
\title{Collaboration in the Sky: A Distributed Framework for Task Offloading and Resource Allocation in Multi-Access Edge Computing}
%
%
%

	
	\author{Yan~Kyaw~Tun,~
	        Tri~Nguyen~Dang,~
	        Kitae~Kim,~
   	        Madyan~Anselwi, 
	    Walid~Saad,~\IEEEmembership{Fellow,~IEEE,~}\\
	and~Choong~Seon~Hong,~\IEEEmembership{Senior~Member,~IEEE}

\thanks{Yan Kyaw Tun,Tri Nguyen Dang, Kitae Kim, Madyan Anselwi, and Choong Seon Hong  are with the Department of Computer Science and Engineering, Kyung Hee University,  Yongin-si, Gyeonggi-do 17104, Rep. of Korea, e-mails:{\{ykyawtun7,trind,  glideslope,  malsenwi, cshong\}@khu.ac.kr}.}
\thanks{Walid Saad is with the Bradley Department of Electrical and Computer Engineering, Virginia Tech, Blacksburg, VA 24061 USA, and also with the Department of
    	Computer Science and Engineering, Kyung Hee University, Yongin-si,
    	Gyeonggi-do 17104,  Rep. of Korea, email: \ {walids@vt.edu}.}}


%
%

\markboth{Journal of \LaTeX\ Class Files,~Vol.~11, No.~4, December~2012}%
{Shell \MakeLowercase{\textit{et al.}}: Multi-Agent Reinforcement Learning Based Resource Sharing in Collaborative UAVs}
%



\maketitle


\begin{abstract}
Recently, unmanned aerial vehicles (UAVs) assisted multi-access edge computing (MEC) systems emerged as a promising solution for providing computation services to mobile users outside of terrestrial infrastructure coverage. As each UAV operates independently, however, it is challenging to meet the computation demands of the mobile users due to the limited computing capacity at the UAV's MEC server as well as the UAV's energy constraint. Therefore, collaboration among UAVs is needed. In this paper, a collaborative multi-UAV-assisted MEC system integrated with a MEC-enabled terrestrial base station (BS) is proposed. Then, the problem of minimizing the total latency experienced by the mobile users in the proposed system is studied by optimizing the offloading decision as well as the allocation of communication and computing resources while satisfying the energy constraints of both mobile users and UAVs. The proposed problem is shown to be a non-convex, mixed-integer nonlinear problem (MINLP) that is intractable. Therefore, the formulated problem is decomposed into three subproblems: i) users tasks offloading decision problem, ii) communication resource allocation problem and iii) UAV-assisted MEC decision problem. Then, the Lagrangian relaxation and alternating direction method of multipliers (ADMM) methods are applied to solve the decomposed problems, alternatively. Simulation results show that the proposed approach reduces the average latency by up to 40.7\% and 4.3\% compared to the greedy and exhaustive search methods.             
\end{abstract}


%
\IEEEpeerreviewmaketitle

\section{Introduction}
\IEEEPARstart{C}{omputing}-intensive applications, such as virtual reality (VR), augmented reality (AR), online gaming, and so on, have become an intrinsic part of our daily activities \cite{chaccour2020can} and \cite{zhou2009virtual}. Nonetheless, data processing on such devices is challenging due to their limited computing capacity and battery life. Thus, a portion of computing tasks can be offloaded from the devices to multi-access edge computing (MEC) servers located at the edge of the networks (e.g., base stations (BSs), access points (APs)) to perform remote computing \cite{mao2017survey} and \cite{mach2017mobile}. As opposed to mobile cloud computing (MCC), MEC can significantly reduce the transmission latency since MEC servers are closer to the devices than cloud servers. However, since radio resources (i.e., bandwidth) are scarce and the MEC servers' computing capacity is limited compared to a cloud, an efficient task offloading, and resource (i.e., communication and computing resources) allocation framework is needed. Therefore, recent research works in MEC such as \cite{you2016energy} and \cite{tun2019energy} have proposed effective computing-intensive task offloading and resource allocation methods.     

 However, installing new terrestrial infrastructures such as BSs, APs, to deploy MEC servers to provide the computing services to the users in temporary events \cite{tun2020energy}, rescue operation in disaster areas, and military operations \cite{samad2007network} will not be cost-effective. Therefore, recently, unmanned aerial vehicles (UAVs) (i.e., balloons, airships, drones) have been deployed as communication and computation platforms for mobile devices due to their flexibility, and cost-effective deployment \cite{mozaffari2019tutorial}. A computation-capable UAV-mounted MEC server can gather data and execute computing tasks of the devices that do not have direct access to the terrestrial base stations (BSs) \cite{cheng2018air}. Moreover, UAV-based MEC servers can also perform as relay stations for the devices located at the edge of the terrestrial BS's coverage area. Although UAV-assisted MEC is a promising technology, several challenges must be addressed before their adoption. First, an efficient task offloading decision is required to minimize the device's and UAV's energy consumption, and the users' latency. Second, due to the scarcity of communication resources (i.e., bandwidth), an effective bandwidth allocation scheme is needed at the UAV. Third, since the available computing resources on a UAV server are limited, we must implement an effective computing resource allocation mechanism on the UAV server. Due to the lack of an effective allocation mechanism, computing energy consumption at the UAV may increase, and the offloaded tasks of devices may not be completed on time. Finally, since each UAV operates independently, it is challenging for the UAV to accomplish the computing tasks of its associated users within the maximum permissible latency of each task (i.e., the task's deadline). As a result, in a UAV-assisted MEC system, collaboration among UAVs can be taken into account for the purpose of reducing the users' latency and increasing energy efficiency.
 
Thus, to address the challenges mentioned above, in this work, a joint optimization problem for a UAV-assisted MEC system including 1) task offloading decision making, 2) communication resource management, 3) computing resource allocation, and 4) collaboration among UAVs is investigated. Further, the objective of the formulated problem is to minimize the total user’s latency, while satisfying the energy budget of both devices and UAVs. To the best of our knowledge, the formulated problem with the corresponding objective has never been investigated in a UAV-assisted MEC system.    

\subsection{Contributions}
The main contribution of this paper is a distributed optimization-based efficient task offloading and resource allocation framework for collaborative multi-UAV-assisted MEC system integrated with a MEC-enabled terrestrial BS. In particular, the key contributions include:  
\begin{itemize}
    \item We first formulate the joint task offloading decision, communication, and computing resource allocation problem in a collaborative multi-UAV-assisted MEC system. The aim of formulated problem is to minimize the latency experienced by the users while satisfying the energy constraint of both users and UAVs. Then, we show that the formulated problem is nonconvex, and mixed-integer nonlinear problem (MILP) that is intractable.  
    
    \item For tractability, we decompose the formulated problem into three subproblems: 1) users tasks offloading decision problem, ii) communication resource allocation problem, and iii) UAV-assisted MEC decision problem. Then, we prove that users tasks offloading problems is a convex problem. Therefore, we use SCS solver in the CVXPY toolkit. 
    
    \item We show that the communication resource allocation problem is convex and can be solved by using a Lagrangian relaxation method.
    
    \item We then show that UAV-assisted MEC decision problem is non-convex. Therefore, to solve the problem, we first relax the binary variables into continuous ones. Then, we apply a distributed algorithm so called the alternating direction method of multipliers (ADMM) approach to address the relaxed problem in a decentralized manner by decoupling the coupling constraint. 
    
    \item Using extensive simulations, we first present the convergence of the proposed algorithm. Then, we compare the average latency experienced by the mobile users under the proposed algorithm with three baselines: centralized algorithm, greedy scheme, and exhaustive search. Simulation results demonstrate that the average latency experienced by the users under our proposed technique is 40.7\% and 4.3\% less than that of the greedy scheme and exhaustive search, respectively. Furthermore, the results also reveal that the average latency experienced by the users under our proposed algorithm is identical to that of the centralized scheme.   
\end{itemize}

The rest of this paper is structured as follows: Section \ref{rel} covers related works. Section \ref{System} describes the system model. The proposed problem formulation and solution approaches are presented in Section \ref{problem} and \ref{solution}. Simulation results are provided in Section \ref{simu}. Finally, conclusions are drawn in Section \ref{con}.

\section{Related Works}
\label{rel}

\subsection{Multi-Access Edge Computing}
We start by an overview on the literature related to multi-access edge computing systems are discussed in this section \cite{yang2018mobile, song2021reward, plachy2021dynamic, yang2020computation, ndikumana2019joint, nduwayezu2020online, li2020joint, zakarya2020epcaware, kiani2017toward, bai2020latency, rivera2020blockchain}. In \cite{yang2018mobile}, the authors proposed a task offloading problem in small cell networks with the objective of minimizing the total energy consumption of all nodes. The work in \cite{song2021reward} introduced a strategy for offloading tasks to MEC servers with the goal of maximizing a reward for each server while operating under a constrained power budget, server computing capacity and wireless network coverage. An efficient communication and computation resource sharing framework in the MEC system is studied in \cite{plachy2021dynamic}. In this work, the authors deployed a probabilistic user mobility modeling to choose the appropriate communication channels between the base station and users and, then, pre-allocate the base station's computation resources. In \cite{yang2020computation}, the authors proposed a multi-task learning-based computing task offloading framework in the MEC system. The work \cite{ndikumana2019joint} formulated joint communication, computation, caching, and control problem in big data MEC systems. Then, the authors applied a block successive upper-bound minimization (BSUM) algorithm in order to address their formulated problem. In \cite{nduwayezu2020online}, the authors introduced a deep reinforcement learning algorithm-based task offloading scheme in a MEC system with non-orthogonal multiple access. The work in \cite{li2020joint} studied the joint optimization of resource allocation and task offloading problem in the MEC environment. In \cite{zakarya2020epcaware}, the authors proposed a game-theoretic approach based cost-effective resource management in the MEC system where they tried to minimize the energy consumption and cost while ensuring the performance of the applications. In \cite{kiani2017toward}, authors presented an auction-based resource allocation problem in hierarchical mobile edge computing, intending to maximize service providers' profits. The work \cite{bai2020latency} focused on a latency minimization problem in the intelligent reflecting surfaces (IRSs) aided multi-access edge computing system. Then, the authors decomposed the proposed problem into multiple subproblems by adopting the block coordinate descent (BCD) technique. Finally, low complexity iterative algorithms are deployed to solve the decomposed subproblems. The blockchain-based secure task offloading scheme in the MEC system was introduced in \cite{rivera2020blockchain}. 

However, all of the studies above assumed that MEC servers are deployed at terrestrial BSs, implying that these studies are fully dependent on terrestrial infrastructures. So, how can you provide communication and computing services to users outside of terrestrial infrastructure coverage, such as those in mountainous areas or deep water, (or) cell edge users experiencing severe link blockage and poor signal strength? Thus, researchers from both academia and industry have recently attempted to employ unmanned aerial vehicles (UAVs) embedded with communication chipsets and computer servers as communication and computation platforms to provide services to the users in the locations mentioned above. 

\subsection{UAVs-Assisted Multi-Access Edge Computing} 
Using a MEC-enabled UAVs have several advantages over a traditional MEC environment as discussed in \cite{wang2021deep, li2020energy, ren2021enabling, chen2020age, xu2021edge, zhan2020completion, liu2020cooperative, wang2020multi, ei2020energy }. For instance, in \cite{wang2021deep}, the authors studied the problem of deep reinforcement learning-based UAV trajectory control in UAV-assisted mobile edge computing. However, the prior work does not take into account task offloading and resource allocation. The work in \cite{li2020energy} studied the problem of joint resource allocation and UAV trajectory optimization in a UAV-assisted MEC system with the aim of maximizing the UAV energy efficiency. Meanwhile, the authors in \cite{ren2021enabling} studied the problem of hierarchical reinforcement learning-based task scheduling in a UAV-assisted mobile edge computing system. In \cite{chen2020age}, the authors proposed a non-cooperative game for efficient resource management scheme with the goal of minimizing the age of information (AoI) in UAV-assisted mobile edge computing. The work in \cite{xu2021edge} introduced a  Stackelberg game-based resource pricing and trading scheme for a blockchain application. The goal of this prior work is to optimally allocate the computing resources at the UAVs and edge computing stations. In \cite{zhan2020completion}, the authors proposed tasks execution time and energy minimization problem in the UAV-enabled MEC system. The joint offloading and efficient resource management scheme was discussed in \cite{liu2020cooperative}. However, all of these prior works \cite{wang2021deep, li2020energy, ren2021enabling, chen2020age, xu2021edge,  zhan2020completion, liu2020cooperative} consider a single UAV. Meanwhile, in \cite{wang2020multi} the authors looked at the multi-UAV case from the perspective of trajectory planning. The goal is to maximize fairness among mobile users in a multi-UAV assisted MEC system. In \cite{ei2020energy}, the authors introduced energy-efficient communication and computation resource management in multi-UAVs-assisted two-stage mobile computation system. Here, the authors hypothesized that a portion of users' offloaded computing tasks are computed by the UAV, while the remaining are forwarded to the MEC-enabled terrestrial BS in order to perform execution. 

The majority of prior works related to UAV-assisted MEC systems focused on a single UAV scenario. Moreover, while some of those works \cite{wang2020multi, ei2020energy} studied a multi-UAV-assisted MEC system, they did not investigate the problem of joint task offloading and communication and computing resources allocation to minimize the total latency experienced by the users. Furthermore, the prior works do not take into account the collaboration among UAVs. In contrast, here, we propose a joint task offloading and resource allocation problem considering collaboration among UAVs in a multi-UAV-assisted MEC system integrated with a MEC-enabled terrestrial BS.

\section{System Model}
\label{System} 
\begin{figure}[t!]
    \centering
    \includegraphics[width=\linewidth]{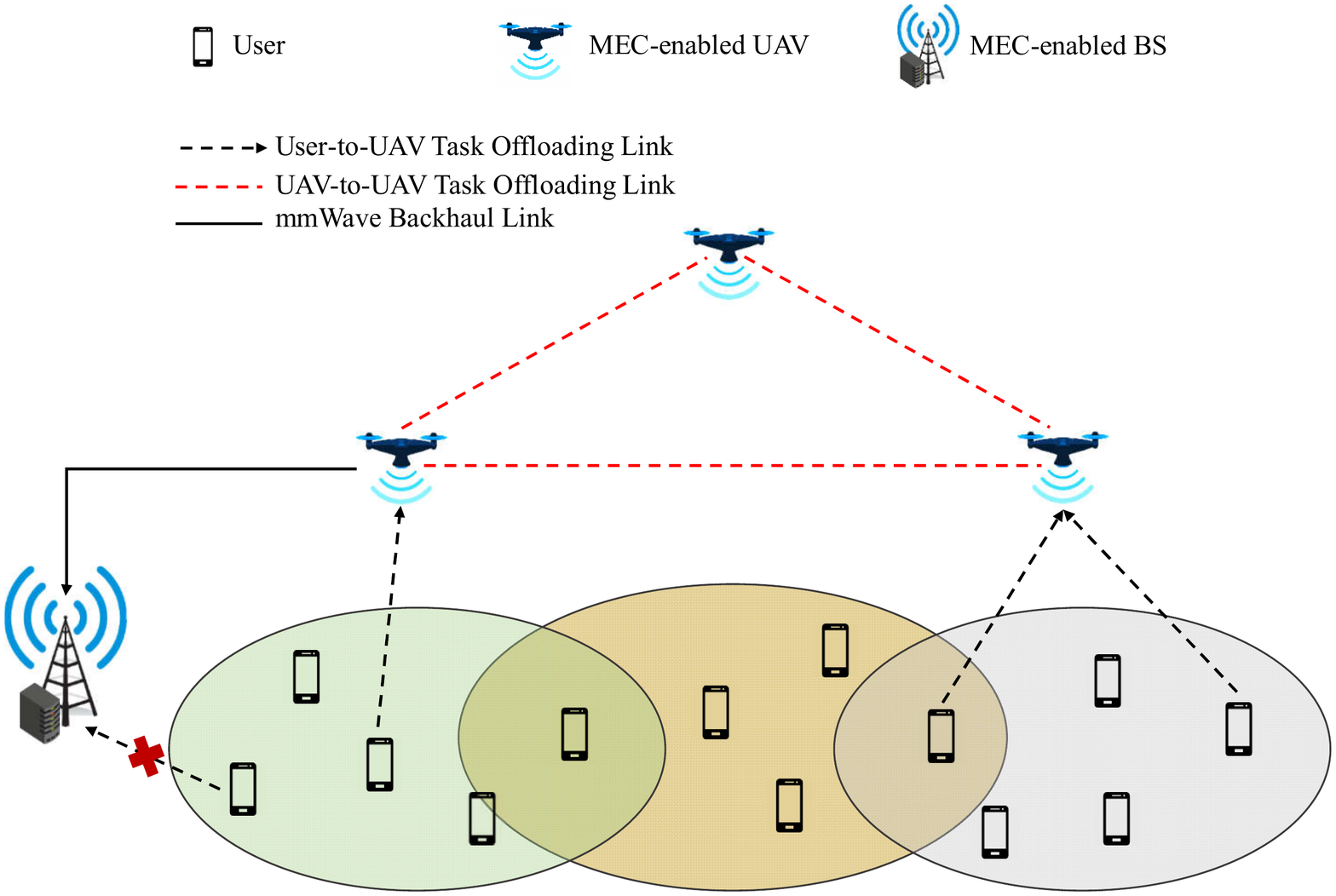}
    \caption{Illustration of our system model.}
    \label{Optimization Framework}
\end{figure}
As illustrated in Fig.~\ref{Optimization Framework}, we consider a multi-access edge computing (MEC) system within an integrated air-ground network consisting of a set $\mathcal{V}$ of $V$ MEC-enabled UAVs and a MEC enabled terrestrial base station (BS). We consider that users are located out of the coverage area of the BS, and, hence, they do not have direct access to the BS. However, users can get wireless access from the nearest UAVs hovering at a certain altitude. For simplicity, we assume that the association of users to serving UAVs is already predetermined depending on the channel quality. In other words, each UAV $v \in \mathcal{V}$ already has its associated set of users, $\mathcal{U}_v$. Therefore, in total, we have a set $ \mathcal{U} = \bigcup\limits_{v=1}^{V} \mathcal{U}_{v} $ of users in the considered network, where $ \mathcal{U}_{v} \cap \mathcal{U}_{w} = \emptyset, \forall v,w\in \mathcal{V}, v \neq w$. Moreover, each user $u \in \mathcal{U}$ has a latency-sensitive computing task $M_u$ that can be expressed by the tuple  $ M_{u} = \big(S_{u}, C_{u}, T_{u}\big)$, where $S_{u}$ is the input data size of the task, $C_{u}$ is the required CPU cycles to compute one bit of data, and $T_u$ is the maximum tolerable latency to compute the task. Due to the limited computing ability of each user's device and the latency constraint of the tasks, it is challenging for the users to compute their tasks locally. Therefore, users can offload a fraction of their tasks to the associated MEC-enabled UAVs via wireless links to perform remote computing. Our proposed network system operates on both 2 GHz and 28 GHz (mmWave) frequency bands. Furthermore, the total available system bandwidth in 2 GHz frequency band is orthogonally divided into two portions: 1) system bandwidth for User-to-UAV data transmission and 2) system bandwidth for UAV-to-UAV data transmission. Therefore, there is no interference between user-to-UAV and UAV-to-UAV data transmission. Meanwhile, the data transmission from UAV-to-BS is taking place on the mmWave band.        

\begin{table*}[htbp]
	\caption{Summary of Notations}
	\renewcommand\arraystretch{1}
	\begin{center}
		\begin{tabular}{|m{1.8cm}|m{6cm}||m{1.8cm}|m{6.3cm}|}
			\hline
			\hfil \textbf{Notation} & \hfil \textbf{Definition} & \hfil \textbf{Notation} & \hfil \textbf{Definition} \\ \hline \hline
			\hfil $\mathcal{V}$ & Set of UAVs, $|\mathcal{V}| = V$ & \hfil $\mathcal{U}$ & Set of users,  $|\mathcal{U}| = U$ \\ \hline
			
			\ \hfil $M_u$ & Computing task of user $u\in \mathcal{U}$ & \hfil  $S_u$ & Total input data size of user $u$'s task \\ \hline			
			\ \hfil  $C_u$ & Required CPU cycles to compute one bit of data & \hfil $T_u$ & Maximum tolerable latency to compute the task $M_u$ \\ \hline			
			\ \hfil $\alpha_u^v$ & Fraction of user $u$'s task offloaded to UAV $v \in \mathcal{V}$& \hfil $(S_u - \alpha_u^v)$ & Fraction of user $u$'s task computed locally\\ \hline
			
			\ \hfil $t_u^{v,\textrm{loc}}$ & Local computation latency experienced by user $u$&  \hfil $f_u$ & Local computing capacity (i.e., cycles/s) of user $u\in \mathcal{U}$ \\ \hline	
			
			\ \hfil $E_u^{v,\textrm{loc}}$ & Local energy consumption of user $u \in \mathcal{U}$ & \hfil $\Gamma_u^v$ & the spectrum efficiency of user $u$ associated with UAV $v$ \\ \hline		
			\ \hfil $P_u$ & Transmit power of user $u$ & \hfil $G_u^v$ & Achievable channel gain between user $u$ and UAV $v$ \\ \hline						
			\ \hfil $\sigma^2$ & Gaussian noise power & \hfil &  \\ \hline					
			\ \hfil $d_u^v$ & Distance between user $u$ and UAV $v$  & \hfil $\varpi$ & Path loass exponent \\ \hline
			\ \hfil $R_u^v$ & Instantaneous achievable uplink tranmission rate of user $u$ &  \hfil $\beta_u^v$ & Fraction of bandwidth allocated to user $u$  \\ \hline			
			\ \hfil $B^v$ & Total available bandwidth at UAV $v$ & \hfil $t_u^{v,\textrm{up}}$ & Uplink transmission latency experienced by user $u$\\ \hline			
			\ \hfil $E_u^{v,\textrm{up}}$ & Energy consumption for uplink transmission & \hfil $\theta_u^{u \rightarrow v}$ & Binary decision variable for indicating whether the offloaded task of user $u$ is computed at the UAV $v$ or not  \\ \hline			
			\ \hfil $t_u^{v,\textrm{comp}}$ & Computation latency at the UAV $v$& \hfil $f_u^v$ & Computing capacity of MEC server at UAV $v$ that is allocated to the user $u$   \\ \hline			
			\ \hfil $F_v^{\mathbf{max}}$ & Maximum computing capacity available at the UAV $v$ & \hfil $t_u^{v,e}$ & Total latency experienced by user $u$ associated with UAV $v$ \\ \hline			
			\ \hfil $E_u^{v,\textrm{comp}}$ & Energy consumption to compute the offloaded task of user $u$ & \hfil $\gamma_u^{v \rightarrow w}$ &  Binary decision variable for indicating whether the offloaded task of user $u$ is forwarded from UAV $v$ to UAV $w$ or not  \\ \hline				
			\ \hfil $R^{v \rightarrow w}$ & Achievable rate between UAV $v$ and UAV $w$ & \hfil $P_v$ & Transmit power of UAV $v$ \\ \hline			
			\ \hfil $G^{v \rightarrow w}$ & Achievable channel gain between UAV $v$ and UAV $w$ &  \hfil  $E^{v \rightarrow w}$ & Transmission energy for forwarding offloaded tasks of users from UAV $v$ to UAV $w$\\ \hline						
			\ \hfil $t_u^{v \rightarrow w,\textrm{com}}$ & Transmission latency between UAV $v$ and UAV $w$ & \hfil $ \phi_u^{v \rightarrow 0}$ & Binary decision variable for indicating whether the offloaded task of user $u$ is forwarded from UAV $v$ to the BS or not \\ \hline					
			\ \hfil $ R^{v \rightarrow 0}$ & Backhaul capacity of the mmWave link between UAV $v$ and the BS & \hfil $B_{\textrm{mm}}^{v \rightarrow 0}$ & The available mmWave backhaul bandwidth between UAV $v$ and the $BS$ \\ \hline			
			\ \hfil  $P^{v \rightarrow 0}$ & Transmit power of the UAV $v$ to the BS & \hfil $G_v^{\textrm{tx}}$ & Antenna gain of the transmitter, UAV $v$\\ \hline  			 	
			\ \hfil $G_0^{\textrm{rx}}$ & Antenna gain of the receiver, BS & \hfil $\bar{P}_{0,v}$ & Received power at the BS  \\ \hline						
			\ \hfil $d_v^0$ & Distance between UAV $v$ and the BS & \hfil $E^{v \rightarrow 0}$ & Transmission energy for forwarding offloaded tasks of users from UAV $v$ to the BS \\ \hline									
			\ \hfil $t_u^{v \rightarrow 0, \textrm{com}}$ & Transmission latency between UAV $v$ and the BS & \hfil $t_u^{v, \textrm{off}}$ & Total latency experienced by the user $u$ when executing a fraction of their task remotely \\ \hline
			\ \hfil $E_v^{\textrm{tot}}$ & Total energy consumption at UAV $v$ & \hfil $E^{v, \textrm{hov}}$ & Hovering energy of UAV $v$ \\ \hline
			
			\ \hfil $t^{v,\textrm{hov}}$ & Hovering time of UAV $v$ & \hfil $E^{v, \textrm{hov}}$ & Hovering energy of UAV $v$ \\ \hline
		    
			\ \hfil $\eta$ & Trust that is proportional to the mass of UAV $v$ & \hfil $\varphi_{v}$ & Power efficiency of UAV $v$ \\ \hline	
			
			\ \hfil $j$ & Number of rotors in each UAV & \hfil $r$ & Diameter of rotor \\ \hline
				
			\ \hfil $\varkappa$ & Air density & \hfil &  \\ \hline
		   
		\end{tabular}
		\label{tab1}
	\end{center}
\end{table*}

\subsection{Local Computing Model}
Let $\alpha_u^v$ be the fraction of task of user $u$ that is offloaded to the associated UAV $v$ for remote computing. Therefore, $(S_u - \alpha_u^v)$ is the fraction of the task computed at the user's device. Thus, the total local computation latency experienced by user $u$ is as follows: 
\begin{equation}
t_u^{v,\textrm{loc}} = \frac{C_u(S_u-\alpha_u^v)}{f_u}, \forall u\in \mathcal{U}, \forall v\in \mathcal{V},   
\end{equation}
where $f_u$ is the computing capacity/computing resource (i.e., cycles/s) of user $u$. Then, the local energy consumption of user $u$ to complete the task execution as follows:
\begin{equation}
    E_{u}^{v,\textrm{loc}} = \kappa (f_{u})^{2}C_u(S_{u} - \alpha_u^v), \forall u \in \mathcal{U}, \forall v \in \mathcal{V}, 
\end{equation}
where $\kappa  = 5 \times 10^{-27}$ is constant which depends on the chip architecture of the user's device.

\subsection{Remote Computing Model}
Each user $u \in \mathcal{U}$ first offloads a fraction of computing task to its associated UAV $v \in \mathcal{V}$ via wireless link. Therefore, we can express the achieving spectrum efficiency between user $u$ and UAV $v$ defined in \cite{zhou2017resource}, as follows:
\begin{equation} 
\Gamma_{u}^{v} = \log_2\left(1 + \frac{P_uG_u^v}{\sigma^2}\right), \forall u \in \mathcal{U}, \forall v \in \mathcal{V},  
\end{equation} 
where $P_u$ is the transmit power of user $u$, $G_u^v$ is the channel gain between user $u$ and UAV $v$, and $\sigma^2$ is the Gaussian noise power. Typically, the channel gain between user $u$ and UAV $v$ can be defined as follows:
\begin{equation}
G_u^v = 10^{-\delta_u^v/10},  \forall u \in \mathcal{U}, \forall v \in \mathcal{V},
\end{equation}  
where $\delta_u^v$ is the path loss between UAV $v$ and user $v$. Furthermore, in this work, we take into account both line-of-sight (LoS) and non-line-of-sight (NLoS) links for the Air-to-ground communication (i.e., users to UAVs communication). Therefore, the path loss $\delta_u^v$ between user $u$ and UAV $v$ is composed of two types of path loss:  LoS path loss, $\delta_u^{v,\textrm{LoS}}$ and non-LoS path loss, $\delta_u^{v,\textrm{LoS}}$. Then, $\delta_u^{v,\textrm{LoS}}$ and $\delta_u^{v,\textrm{LoS}}$ defined in \cite{mozaffari2016efficient} as follows:
\begin{equation}
 \delta_u^{v,\textrm{LoS}} = 2n\log\bigg(\frac{4\pi d_u^vB_c^{\textrm{lte}}}{c}\bigg) + L_{\textrm{LoS}}, 
\end{equation}
\begin{equation}
\delta_u^{v,\textrm{NLoS}} = 2n\log\bigg(\frac{4\pi d_u^vB_c^{\textrm{lte}}}{c}\bigg) + L_{\textrm{NLoS}}, 
\end{equation}         
where $n \geq 2$ is the path loss exponent, $B_c^{\textrm{lte}}$ represents the carrier frequency (i.e., 2 GHz), $c$ is the speed of light, $ L_{\textrm{LoS}}$ and $ L_{\textrm{NLoS}}$ are the average added losses for the LoS and NLoS links. Furthermore, $d_u^v$ is the distance between user $u$ and UAV $v$, and it can be expressed as follows: 
\begin{equation}
d_u^v= \sqrt{ (x_v - x_u)^2 + (y_v - y_u)^2 + h_v^2}, \forall u \in \mathcal{U}, \forall v \in \mathcal{V}. 
\end{equation}

Then, the probability of existing LoS connectivity between UAV $v$ and user $u$ defined in \cite{mozaffari2016efficient}\cite{manzoor2021ruin} as follows:   
\begin{equation} 
\textrm {Pr}_u^{v,\textrm{LoS}} = \frac {1}{1+ C~\textrm {exp}\left[{D\left({\frac {180}{\pi } \tan ^{-1} \frac {h_{u}}{d_{uk}} } - C \right)}\right]},
\end{equation}
where $h_v$ is the hovering altitude of UAV $v$, and C and D are constants which depend on the environment, such as urban, rural, suburban, and so on. Consequently, the probability of existing NLoS link between user $u$ and UAV $v$ can be calculated as follows:
\begin{equation}
\textrm {Pr}_u^{v,\textrm{NLoS}} = 1 - \textrm {Pr}_u^{v,\textrm{LoS}}.
\end{equation} 
Therefore, the total path loss between user $u$ and UAV $v$ is given as follows:
\begin{equation}
 \delta_u^v = \textrm {Pr}_u^{v,\textrm{LoS}}\delta_u^{v,\textrm{LoS}} +  {Pr}_u^{v,\textrm{NLoS}}\delta_u^{v,\textrm{NLoS}}, \forall u \in \mathcal{U}, \forall v \in \mathcal{V}. 
\end{equation}
        
Finally, the instantaneous achievable uplink transmission rate of user $u$ is as follows:
\begin{equation}
   R_u^v = \beta_u^v B^v \Gamma_u^v, \forall u \in \mathcal{U}, \forall v \in \mathcal{V}, 
\end{equation}
where $\beta_u^v$ is the fraction of bandwidth allocated to offloading user $u$, and $B^v$ is the total available bandwidth at UAV $v$.

Therefore, depending on the uplink transmission rate, we formulate the uplink transmission latency experienced by user $u$ as follows:
\begin{equation}
    t_u^{v,\textrm{up}} = \frac{\alpha_u^v}{R_u^v}, \forall u \in \mathcal{U}_v, \forall v \in \mathcal{V}. 
\end{equation} 
Then, the energy consumption for uplink transmission from user $u$ to UAV $v$ as follows:
\begin{equation}
    E_u^{v, \textrm{up}} = \frac{P_u\alpha_u^v}{R_u^v}, \forall u \in \mathcal{U}_v, \forall v \in \mathcal{V}, 
\end{equation}

In our model, the offloaded task of user $u \in \mathcal{U}$ is executed on the MEC server of the associated UAV $v \in \mathcal{V}$ if it has enough computing capacity. Therefore, we define the binary variable $\theta_u^{u \rightarrow v} \in \{0,1\}$ as a decision variable, which captures whether or not the offloaded task of user $u$ is computed at the UAV $v$, where $\theta_u^{u \rightarrow v}$ is given by:
\begin{equation}
\theta_{u}^{u \rightarrow v} = 
\begin{cases}
1, \ \ \text{it the offloaded task of user $u$ is computed}\\
\ \ \ \ \ \text{ at UAV $v$}, \\
0, \ \ \text{otherwise}.
\end{cases}
\end{equation}

Therefore, the computation latency when the offloaded task of user $u$ is computed at UAV $v$, i.e., $\theta_{u}^{u \rightarrow v} =1$, is as follows: 
\begin{equation}
   t_u^{v, \textrm{comp}} = \frac{C_u\alpha_u^v}{f_u^v}, \forall u\in \mathcal{U}_v, \forall v \in \mathcal{V},
\end{equation}
where $f_u^v$ is the computing capacity (i.e., cycles/s) of the MEC server of the UAV $v$ that is allocated to compute the task of user $u$ and it can be calculated based on the proportional allocation defined in \cite{tun2019wireless}, as follows: 
\begin{equation}
    f_u^v = \frac{\alpha_u^v}{ \sum\limits_{q \in  \mathcal{U}_v}  \alpha_q^v} F_v^{\mathbf{max}}, \forall u \in \mathcal{U}_v, \forall v \in \mathcal{V}, 
\end{equation}
where $F_v^{\mathbf{max}}$ is the maximum computing capacity available at the UAV $v$. Here, the following constraint must be satisfied when allocating the computing capacity of each UAV $v \in \mathcal{V}$ to its associated users:
\begin{equation}
    \sum\limits_{ u \in \mathcal {U}_v} \theta_u^{u \rightarrow v} f_u^v \leq  F_v^{\mathbf{max}},   \forall v \in \mathcal{V}.
\end{equation}
Then, the latency experienced by the user $ u \in \mathcal{U} $ in order to complete the execution of its offloaded computing task at UAV $v \in \mathcal{V}$ can be expressed as follows:
\begin{equation}
    t_u^{v,e} =   t_u^{v,\textrm{up}} +   t_u^{v, \textrm{comp}}, \ \ \forall u \in \mathcal{U}_v, \forall v \in \mathcal{V}. 
\end{equation}

Finally, the energy consumption needed to compute the offloaded task of user $u$ is given by: 
\begin{equation}
    E_u^{v, \textrm{comp}} = \kappa_v (f_u^v)^2C_u\alpha_u,  \forall v \in \mathcal{V},    
\end{equation}
where $k_v = 5 \times 10^{-27}$ is a constant which depends on the chip architecture of the server at the UAV.

When the computing capacity at UAV $v$ is not sufficient in order to compute the offloaded tasks of its associated users, the UAV $v$ forwards the tasks to the neighboring UAVs $w \in \mathcal{V}, v \neq w$ via wireless link. Therefore, we define the binary variable $\gamma_u^{v \rightarrow w} \in \{0,1\}$ as a decision variable, which captures whether or not the offloaded computing task of user $u$ is forwarded from the UAV $v$ to a neighboring UAV $w$, as follows:
\begin{equation}
\gamma_{u}^{v \rightarrow w} = 
\begin{cases}
1, \ \ \text{if offloaded computing task of user $u$ is}\\
\ \ \ \ \ \text{forwarded from UAV $v$ to UAV $w$}, \\
0, \ \ \text{otherwise}.
\end{cases}
\end{equation}
Then, the offloading (i.e., transmission) latency between UAV $v$ and UAV $w$ can be given by:
\begin{equation}
    t_u^{v \rightarrow w,\textrm{com}} =  \frac{\sum\limits_{u \in \mathcal{U}_v} \gamma_{u}^{v \rightarrow w}\alpha_u^v}{R^{v\rightarrow w}},  \forall v,w\in \mathcal{V}, v \neq w,
\end{equation}
where $R_u^{v\rightarrow w}$ is the achievable rate between UAV $v$ and UAV $w$ defined as:
\begin{equation}
  R^{v\rightarrow w}= B^{v\rightarrow w} \log_2\left(1+ \frac{P_vG^{v\rightarrow w}}{\sigma^2}\right),  \forall v,w \in \mathcal{V}, v \neq w,  
\end{equation}
where $B^{v\rightarrow w}$ is the available bandwidth for direct communication between UAV $v$ and UAV $w$, $P_v$ is the transmit power of UAV $v$, and $G^{v\rightarrow w}$ is the achievable channel gain between UAVs. Here, we consider line-of-sight (LoS) link for UAV-to-UAV communication. Therefore, the channel gain between UAVs which is defined in \cite{challita2017network}, as follows:    
\begin{equation}
    G^{v\rightarrow w} = 10^{-\left(L^{v\rightarrow w}/10\right)}, \forall v,w\in \mathcal{V}, 
\end{equation}
where $L^{v \rightarrow w} = \Theta^{v\rightarrow w} + \Upsilon_{\textrm{LoS}} $ is the path loss between UAV $v$ and $v$. Here, $\Upsilon_{\textrm{LoS}}$ is the additional attenuation factor for LoS link \cite{challita2017network}, and $\Theta^{v\rightarrow w}$ is as follows:
\begin{equation}
\Theta^{v\rightarrow w} (dB)= 20 \log_{10}(d_v^w) + 20\log_{10}(f_c) + 10\log_{10}\left[\left(\frac{2\pi}{c}\right)^2\right],  
\end{equation}
where, $c$ is the speed of light, $f_c$ is the carrier frequency (i.e., 2 GHz). Furthermore, $d_v^w$ is the distance between UAV $v$ and UAV $w$, and it can be expressed as follows: 
\begin{equation}
    d_v^w = \sqrt{ (x_w - x_v)^2 + (y_w - y_v)^2 + (h_v -h_w)^2}, \forall v,w \in \mathcal{V}, 
\end{equation}
where $[x_w, y_w]^T$ and $h_w$ are the horizontal coordinate and the hovering altitude of UAV $w$. Then, the offloading (i.e., transmission) energy as follows:  
\begin{equation}
    E^{v \rightarrow w} = P_v \left(\frac{\sum\limits_{u\in \mathcal{U}_v} \gamma_u^{v \rightarrow w}\alpha_u^v}{R^{v \rightarrow w}}\right),  \forall v,w \in \mathcal{V},
\end{equation}
Here, $t_u^{v \rightarrow w,\textrm{comp}}$ and $E_u^{v \rightarrow w,\textrm{comp}}$ are the latency and energy consumption when the task of $ u \in \mathcal{U}_v$ is computed on the MEC server of UAV $w \in \mathcal{V}, v \neq w$ and which can be calculated based on (15) and (19). Therefore, the total latency when the offloaded task of user $u$ associated with UAV $v$ is computed at the UAV $w$ is as follows: 
\begin{equation}
    t_u^{v\rightarrow w,e} =   t_u^{v,\textrm{up}} + t_u^{v \rightarrow w, \textrm{com}}+  t_u^{v \rightarrow w,\textrm{comp}}, \ \ \forall u \in \mathcal{U}_v, \forall v,w \in \mathcal{V}. 
\end{equation}

When the computing capacity at neighboring UAVs are not sufficient, then, the UAV $v$ forwards the computing task of its associated user $u$ to the terrestrial BS. Thus, we now define the binary variable $\phi_u^{v \rightarrow 0}$ as the decision variable, which captures whether or not the computing task of user $u$ is forwarded to the terrestrial base station through the mmWave (i.e., Ka band 28 GHz), as follows:
\begin{equation}
\phi_{u}^{v \rightarrow 0} = 
\begin{cases}
1, \ \ \text{if offloaded computing task of user $u$ }\\
\ \ \ \ \ \text{is forwarded from UAV $v$ to the BS}, \\
0, \ \ \text{otherwise}.
\end{cases}
\end{equation}
Next, the transmission latency between UAV $v$ and the terrestrial base station when the computing task of user $u$ is forwarded to the BS, i.e., $\gamma_{u}^{v \rightarrow 0}=1$, will be given by:
\begin{equation}
    t_u^{v \rightarrow 0,\textrm{com}} =  \frac{\sum\limits_{u \in \mathcal{U}_v} \phi_{u}^{v \rightarrow 0}\alpha_u^v}{R^{v\rightarrow 0}},  \forall v\in \mathcal{V},
\end{equation}
where $R^{v\rightarrow 0}$ is the backhaul capacity of the mmWave link between UAV $v$ and terrestrial base station, which is given by \cite{lai2019data}:
\begin{equation}
 R^{v\rightarrow 0} = B_{\textrm{mm}}^{v\rightarrow 0} \log_2\left(1 + \frac{\bar{P}_{0,v}}{B_{\textrm{mm}}^{v\rightarrow 0} \sigma^2}\right), \forall v \in \mathcal{V}, 
\end{equation}

Here, $B_{\textrm{mm}}^{v\rightarrow 0}$ is the available mmWave backhaul bandwidth between UAV $v$ and BS, $\bar{P}_{0,v}$ is the received power at the BS from the UAV $v$ and it is given by:
\begin{equation}
    \bar{P}_{0,v} = P^{v \rightarrow 0} G_v^{\textrm{tx}} G_0^{\textrm{rx}} \left( \frac{c}{4\pi d_v^0 B_c^{\textrm{mm}}}\right), 
\end{equation}
where $P^{v \rightarrow 0}$ is the transmit power of UAV $v$, $c$ is the speed of light, and $B_c^{\textrm{mm}}$ is the carrier frequency of the mmWave backhaul link. $ G_v^{\textrm{tx}}$ and $G_0^{\textrm{rx}}$ are the antenna gains of the transmitter, UAV $v$, and the receiving BS. Finally, $d_v^0$ is the distance between UAV $v$ and the BS, given by:
\begin{equation}
    d_v^0 = \sqrt{ (x_0 - x_v)^2 + (y_0 - y_v)^2 + h_v^2}, \forall v \in \mathcal{V}, 
\end{equation}
where $[x_0, y_0]^T$ is the location of the terrestrial BS. Then, the transmission energy of of the UAV $v$ will be:
\begin{equation}
    E^{v \rightarrow 0} = P^{v \rightarrow 0} \left( \frac{\sum\limits_{u\in \mathcal{U}_v} \phi_u^{v \rightarrow 0} \alpha_u^v}{R^{v \rightarrow 0}}\right),  \forall v \in \mathcal{V}. 
\end{equation}

Thus, the offloaded task from $u \in \mathcal{U}_v$ to MEC server located at BS exhibit $t_u^{v \rightarrow 0, \mathrm{comp}}$ latency  and $E_u^{v \rightarrow 0, \mathrm{comp}}$ energy consumption which can be computed as (15) and (19). However, since the BS is connected to the grid, the energy consumption constraint at the BS is not as stringent as the energy constraint faced by users and UAVs. Thus, we have omitted the computation energy consumption at the BS. The total latency when the offloaded task of user $u$ associated with UAV $v$ is computed at the BS will be given by:
\begin{equation}
    t_u^{v\rightarrow 0,e} =   t_u^{v,\textrm{up}} + t_u^{v \rightarrow 0, \textrm{com}}+  t_u^{v \rightarrow 0,\textrm{comp}}, \ \ \forall u \in \mathcal{U}_v, \forall v \in \mathcal{V}. 
\end{equation}
Furthermore, we next define $t_u^{v,\textrm{off}}$, as the total transmission and computation latency experienced by user $u \in \mathcal{U}_v$ to execute fraction of its computing task remotely, as follows: 
\begin{equation}
\begin{split}
    t_u^{v,\textrm{off}} = \theta_u^{u \rightarrow v} t_u^{v,e} + \sum\limits_{w \in \mathcal{V}, w \neq v}  \gamma_u^{v \rightarrow w} t_u^{v \rightarrow w,e} + \phi_u^{v \rightarrow 0} t_u^{v \rightarrow 0, e}, \\
    \forall u\in \mathcal{U}_v, \forall v \in \mathcal{V}. 
\end{split}
\end{equation}
Here, $E_v^{\textrm{tot}}$ is the total energy consumption of UAV $v$ as, which includes, i) the computation energy for its associated users, ii) the transmission energy of UAV $v$ to neighboring UAVs, iii) the transmission energy of UAV $v$ to the BS, and iv) the hovering energy, as follows:
\begin{equation}
\begin{split}
    E_v^{\textrm{tot}} = \sum\limits_{ u \in \mathcal{U}_v} \theta_u^{u \rightarrow v}  E_u^{v, \textrm{comp}} + \sum\limits_{ w \in \mathcal{V}, w \neq v}  E^{v \rightarrow w} + E^{v \rightarrow 0} \\
    + E^{v, \text{hov}},  \ \ \forall v \in \mathcal{V}, 
\end{split}
\end{equation}
where $E^{v, \text{hov}}$ is the hovering energy and given by \cite{monwar2018optimized}:
\begin{equation}
\begin{split}
    E^{v,\textrm{hov}} &= P^{v, \textrm{hov}} t^{v, \textrm{hov}}, \forall v \in \mathcal{V}, \\
    &= \frac{\eta \sqrt{\eta}}{\varphi_v\sqrt{0.5\pi jr^2\varkappa}} t^{v,\textrm{hov}}, \forall v \in \mathcal{V}. 
\end{split} 
\end{equation}
where $\eta$ is the trust that is proportional to the UAV's mass, $\varphi_v$ is the power efficiency of the UAV $v$, $j$ is the number of rotors in each UAV, $r$ is the diameter of each rotor, and $\varkappa$ is the air density. Finally, $t^{v, \textrm{hov}}$ is the maximum hovering time of UAV, and it is given by:
\begin{equation}
\begin{split}
    t^{v,\textrm{hov}} = \max\limits_{u \in \mathcal{U}_v} \Bigg\{t_u^{v,\textrm{up}} + \max\bigg(t_u^{v,\textrm{comp}},  t_u^{v \rightarrow w, \textrm{com}}+  t_u^{v \rightarrow,  w,\textrm{comp}},  \\
     t_u^{v \rightarrow 0, \textrm{com}}+  t_u^{v \rightarrow 0,\textrm{comp}}\bigg) \bigg\}
    \end{split}
\end{equation}
\vspace{-0.4in} 
\section{Problem Formulation}
\label{problem} 
\begin{figure*}[t]
\begin{equation}
\begin{aligned}
     \mathbf{Z}_{v}(\boldsymbol{\alpha, \beta, \theta, \gamma, \phi} ) &=  \left(\sum\limits_{u \in U_{v}} t_u^{v,\textrm{loc}} + \theta_u^{u \rightarrow v} t_u^{v,e} + \sum\limits_{w \in \mathcal{V}, w \neq v}  \gamma_u^{v \rightarrow w} t_u^{v \rightarrow w,e} + \phi_u^{v \rightarrow 0} t_u^{v \rightarrow 0, e}\right),  \forall v \in \mathcal{V}, 
\end{aligned}
\end{equation}
\end{figure*}

In this paper, a collaborative multi-UAV-assisted MEC system integrated with a MEC-enabled terrestrial BS is considered. Then, for the considered system given in \ref{System}, we investigate the problem of joint computing task offloading and resource allocation, aiming at minimizing users' latency.
Therefore, the objective function of the proposed latency minimization problem is defined as $\mathbf{Z}(\boldsymbol{\alpha, \beta, \theta, \gamma, \phi}) $ = $ \sum\limits_{v \in \mathcal{V}}\mathbf{Z}_{v}(  \boldsymbol{\alpha, \beta, \theta, \gamma, \phi} )$. 

To our knowledge, this is the first study that investigates the latency minimization problem in collaborative multi-UAV-assisted MEC system, by jointly optimizing offloading decision, communication and computing resource allocation. The problem can be formally posed as follows:
	\begin{mini!}[2]                 
		{\boldsymbol{\alpha, \beta, \theta, \gamma, \phi}}                               
		{\mathbf{Z}(\boldsymbol{\alpha, \beta, \theta,\gamma, \phi})}{\label{opt:P1}}{\textbf{P:}} 
		\addConstraint{ E_{u}^{\textrm{loc}}+ E_u^{v,\textrm{up}} \leq E_u^{\mathbf{max}}, \ \ \ \ \ \ \forall u \in \mathcal{U}_v,}
		\addConstraint{ \sum\limits_{ u \in \mathcal{U}_v} \theta_u^{u \rightarrow v}  E_u^{v, \textrm{comp}} + \sum\limits_{ w \in \mathcal{V}, w \neq v}  E^{v \rightarrow w}}\nonumber
		\addConstraint{ + E^{v \rightarrow 0} + E^{v, \text{hov}} \leq E_v^{\mathbf{max}},  \forall v \in \mathcal{V},}
		\addConstraint{\sum\limits_{u \in \mathcal{U}_{v} } E_{u}^{v \rightarrow w, \textrm{comp}} + E^{w,\textrm{hov}} \leq E_{w}^{\mathbf{max}},} \nonumber
		\addConstraint{ \ \ \ \ \ \ \ \ \ \ \ \ \ \ \ \ \ \ \ \ \ \ \forall w \in \mathcal{V}, \  w \neq v} 
		\addConstraint{ \sum_{u\in \mathcal{U}_v} \beta_u^v \leq 1,  \ \ \ \  \forall v \in \mathcal{V},}
		\addConstraint{ \beta_u^v \in [0,1],  \ \ \ \ \ \  \forall u \in \mathcal{U}_v, \forall v \in \mathcal{V},}
		\addConstraint{\theta_u^{u \rightarrow v} + \sum_{\substack{w\in \mathcal{V}, \\ w\neq v}} \gamma_u^{v \rightarrow w} + \phi_u^{v \rightarrow 0} = 1, \forall u \in \mathcal{U}_v, } 
		\addConstraint{ 0 \leq \alpha_u^ v\leq S_u, \forall u \in \mathcal{U}, \forall v \in \mathcal{V},}
		\addConstraint{\theta_u^{ u\rightarrow v} \in \{0,1\}, \forall u \in \mathcal{U}_u, \forall v \in \mathcal{V},}
		\addConstraint{\gamma_u^{v \rightarrow w} \in \{0,1\}, \forall u \in \mathcal{U}_u, \forall v,w \in \mathcal{V},}
		\addConstraint{\phi_u^{v \rightarrow 0} \in \{0,1\}, \forall u \in \mathcal{U}_u, \forall v \in \mathcal{V},}
	\end{mini!}

where $\boldsymbol{\alpha}$ is users' tasks offloading decision vector with each element $\alpha_u^v$ representing the offloaded task of user $ u \in \mathcal{U}_v$ associated with UAV $v \in \mathcal{V}$, and $\boldsymbol{\beta}$ is the communication resource (i.e., bandwidth) allocation vector with each element $\beta_u^v$ representing the fraction of bandwidth allocated to user $u\in \mathcal{U}_v$ at UAV $v \in \mathcal{V}$. $\boldsymbol{\theta}$ is the computing decision vector of UAV-assisted MEC with each element $\theta_u^{u \rightarrow v}$ indicating the offloaded task of user $u \in \mathcal{U}_v$ which is computed at UAV $v$ or not. Moreover, $\boldsymbol{\gamma}$ and $\boldsymbol{\phi}$ are tasks offloading decision vectors of UAV-assisted MEC. (40b), (40c), and (40d) represent the energy constraint of both mobile users and UAVs. (40e) and (40f) constrain the total communication resource (i.e., bandwidth) allocated to all users to be smaller than the maximum available bandwidth at each associated UAV. The constraint in (40g) guarantees that the offloaded tasks of users have to be computed at only one position, i.e., at the associated UAV (or) at the neighboring UAVs (or) at the BS. The constraint in (40h) shows that the fraction of the task of user $u$ offloaded to its associated UAV $v$ is always less the total data size of the task. Finally, (40i), (40j), and (40k) show the binary decision variables.

As a result of the coupling between decision variables (i.e., $\boldsymbol{\alpha}$, $\boldsymbol{\beta}$, $\boldsymbol{\theta}$, $\boldsymbol{\gamma}$, and $\boldsymbol{\phi}$) in both objective function and constraints, the nonlinear constraints, mixture of continuous and binary variables, and non-convex structure, the formulated optimization problem in (40) is challenging to solve. Moreover, as the formulated problem involves in a combinatorial category, it is intractable for large scale setting. Thus, it is unsolvable in practical setting. Therefore, first by adopting the block coordinate descent (BCD) technique, the formulated problem is decomposed into three tractable subproblems. Then, the decomposed subproblems are transformed into convex problems and solved alternatively. Due to convexity of each subproblem, we can achieve the optimal solution of the equivalent problem but not the original problem.        
\vspace{-0.4cm}
\section{Proposed Solution}
\vspace{-0.3cm}
\label{solution} 
\subsection{Users Tasks Offloading Decision (UTOD)}
For a given bandwidth allocation $\boldsymbol{\beta}$, and UAV-assisted MEC decision variables, i.e., $\boldsymbol{\theta}, \boldsymbol{\gamma}$, and $\boldsymbol{\phi}$, the users tasks offloading decision problem can be formulated as follows:
\vspace{-0.1cm}
	\begin{mini!}[2]                 
		{\boldsymbol{\alpha}}                               
		{\mathbf{Z}(\boldsymbol{\alpha})}{\label{opt:P2}}{\textbf{P1:}} 
		\addConstraint{ E_{u}^{\textrm{loc}}+ E_u^{v,\textrm{up}} \leq E_u^{\mathbf{max}}, \ \ \ \ \ \ \forall u \in \mathcal{U}_v,}
		\addConstraint{ 0 \leq \alpha_u^ v\leq S_u, \forall u \in \mathcal{U}, \forall v \in \mathcal{V},}
		\vspace{-0.2cm}
  \end{mini!}
\vspace{-0.1cm} 
\textbf{Lemma 1.} \textit{At the given bandwidth allocation and UAV-assisted MEC decision, the optimization problem in (\ref{opt:P2}) is a convex problem.} 
\begin{IEEEproof}
The first order derivative of the objective function in (41a) w.r.t to $\alpha_u^v$ can be expressed as follows:  
\begin{equation}
 \frac{\partial \mathbf{Z}(\boldsymbol{\alpha})}{\partial \alpha_u^v} = - C_u + \frac{1}{R_u^v}, \forall u \in \mathcal{U}_v, \forall v \in \mathcal{V}.
\end{equation}

Then,
\begin{equation}
    \frac{\partial^2 \mathbf{Z}(\boldsymbol{\alpha})}{\partial (\alpha_u^v)^2} = 0
\end{equation}
From (43), we can conclude that the objective function is a linear function. Moreover, the constraints in (41b) and (41c) are also linear. Therefore, the problem in (41) is convex.
\end{IEEEproof} 
Since the users tasks offloading decision \textit{(UTOD)} problem in (\ref{opt:P2}) is a convex problem, we can solve the problem in (41) by using the CVXPY toolkit. 

\subsection{Communication Resource Allocation (CRA)}
For a given tasks offloading decision $\boldsymbol{\alpha}$, and UAV-assisted MEC decision variables, i.e., $\boldsymbol{\theta}, \boldsymbol{\gamma}$, and $\boldsymbol{\phi}$, communication resource allocation can be formulated as follows :
	\begin{mini!}[2]                 
		{\boldsymbol{\beta}}                               
		{\mathbf{Z}(\boldsymbol{\beta})}{\label{opt:P3}}{\textbf{P2:}} 
		\addConstraint{E_u^{v,\textrm{up}} \leq E_u^{\mathbf{max}}, \ \ \ \ \ \ \forall u \in \mathcal{U}_v,}
		\addConstraint{ \sum_{u\in \mathcal{U}_v} \beta_u^v \leq 1,  \ \ \ \  \forall v \in \mathcal{V},}
		\addConstraint{ \beta_u^v \in [0,1],  \ \ \ \ \ \  \forall u \in \mathcal{U}_v, \forall v \in \mathcal{V},}
  \end{mini!}

\textbf{Lemma 2.} \textit{The communication resource allocation (CRA) problem in (\ref{opt:P3}) is a convex problem.}


\begin{IEEEproof}
  The first order derivative of the objective function in (44a) w.r.t to $\beta_u^v$ is as follows:
\begin{equation}
 \frac{\partial \mathbf{Z}(\boldsymbol{\beta})}{\partial \beta_u^v} = \frac{-\alpha_u^v(\beta_u^v)^{-2}}{B^v \log_2\left(1 + \frac{P_uG_u^v}{\sigma^2} \right) }, \forall u \in \mathcal{U}_v, \forall v \in \mathcal{V}.
\end{equation}

Then,
\begin{equation}
    \frac{\partial^2 \mathbf{Z}(\boldsymbol{\beta})}{\partial (\beta_u^v)^2} = \frac{2\alpha_u^v(\beta_u^v)^{-3}}{B^v \log_2\left(1 + \frac{P_uG_u^v}{\sigma^2} \right) }, \forall u \in \mathcal{U}_v, \forall v \in \mathcal{V}.
\end{equation}
From (46), we can observe that $\frac{\partial^2 \mathbf{Z}(\boldsymbol{\beta})}{\partial (\beta_u^v)^2}>0$. Therefore, we conclude that the objective function of the communication resource allocation problem is convex. Moreover, the constraint in (44b) and (44c) are convex and linear. Thus, the formulated communication resource allocation problem in (44) is convex.    
\end{IEEEproof}

For the constraints in (44b) and (44c), we introduce non-negative Lagrangian multipliers $\zeta_u^v$, and $\xi^v$, respectively. The Lagrangian function of (44) can then be expressed by integrating the objective function in (44a) and constraints (44b), (44c). We can formally posed the Lagrangian function as follows:
\begin{equation}
\begin{split}
\mathcal{L}(\boldsymbol{\beta}, \boldsymbol{\zeta}, \boldsymbol{\xi}) &= \sum\limits_{v \in \mathcal{V}} \sum\limits_{u \in \mathcal{U}_v}\left( \frac{\alpha_u^v}{\beta_u^v B^v \log_2\left(1 + \frac{P_uG_u^v}{\sigma^2} \right)}\right) + \sum\limits_{v\in \mathcal{V}}  \\
 &  \sum\limits_{u\in \mathcal{U}_v} \zeta_u^v(E_u^{v,\textrm{up}} - E_u^{\mathbf{max}}) + \sum\limits_{v\in \mathcal{V}} \xi^v \left(\sum\limits_{u \in \mathcal{U}_v} \beta_u^v -1 \right)
\end{split} 
\end{equation} 

Then, the dual problem of \textbf{P2} is expressed as follows:
\begin{align}
\underset{(\boldsymbol{\zeta} \geq 0, \ \boldsymbol{\xi} \geq 0)}{\max} \
& D( \boldsymbol{\zeta}, \boldsymbol{\xi}),  
\end{align}
where
\begin{equation}
\begin{split}
   D(\boldsymbol{\zeta}, \boldsymbol{\xi}) = \ &  \underset{\boldsymbol{}}{\min} \ \mathcal{L}(\boldsymbol{\beta}, \boldsymbol{\zeta}, \boldsymbol{\xi}) \\
   & \text{subject to} \ \ \text{(44b)},\ \ \text{(44c),  \text{and} \ \text{(44d)}}.
\end{split}
\end{equation}
The communication resource allocation problem in (44) is a convex problem, as seen above, so there is a strictly feasible point where Slater's condition holds, resulting in strong duality \cite{boyd2004convex}. As a result, we can solve the problem in (44) by using the dual problem in (48). The dual problem in (48) can be solved by using the sub-gradient approach, in which the dual variables are updated as follows:
\begin{equation}
    \zeta_u^v(i+1) = \bigg[ \zeta_u^v(i) + \varrho_1(i) \bigg( E_u^{v,\textrm{up}} - E_u^{\mathbf{max}}\bigg) \bigg]^{+}
\end{equation}
\begin{equation}
    \xi^v(i+1) = \bigg[ \xi^v(t) + \varrho_2(i) \left( \sum\limits_{u \in \mathcal{U}_v} \beta_u^v - 1 \right) \bigg]^+
\end{equation}
where $\varrho_l(i) (l =1,2.)$ is a step size and it is given by:
\begin{equation}
    \varrho_l(i) = \frac{m}{\sqrt{i}}, m >0, l = 1,2. 
\end{equation}
  
\textbf{Proportion 1}. \textit{According to the Karush-Kuhn-Tucketer (KKT) conditions \cite{boyd2004convex}, the optimal solution of communication resource allocation problem, \textbf{P2} is given by: }
\begin{equation}
    \beta_u^{v*} = \sqrt{ \frac{\alpha_u^v(1+\zeta_u^vP_u)}{\xi^vB^v \log_2\left(1+ \frac{P_uG_u^v}{\sigma^2} \right)} }
\end{equation}

\begin{IEEEproof}
The first order derivative of the Lagrangian function in (47) w.r.t $\beta_u^v$ is given by:
 \begin{equation}
\begin{split}
\frac{\partial \mathcal{L}(\boldsymbol{\beta}, \boldsymbol{\zeta}, \boldsymbol{\xi})}{\partial \beta_u^v} &= \left[ \frac{-\alpha_u^v}{(\beta_u^v)^2 B^v \log_2\left(1+ \frac{P_uG_u^v}{\sigma^2} \right) }  \right] + \xi^v - \zeta_u^v \\  & \left[\frac{P_u \alpha_u^v}{(\beta_u^v)^2 B^v \log_2\left(1+ \frac{P_uG_u^v}{\sigma^2} \right)}      \right]
 \leq 0,\ \text{if} \ \beta_u^v \geq 0, \\ &\forall u \in \mathcal{U}_v, \forall v \in \mathcal{V},  
\end{split} 
\end{equation} 
Meanwhile $\beta_u^v > 0$, $\frac{\partial \mathcal{L}(\boldsymbol{\beta}, \boldsymbol{\zeta}, \boldsymbol{\xi})}{\partial \beta_u^v}=0$. Thus,
\begin{equation}
    \left[ \frac{-\alpha_u^v(1+ \zeta_u^vP_u)}{(\beta_u^v)^2 B^v \log_2\left(1+ \frac{P_uG_u^v}{\sigma^2} \right) } \right] + \xi^v = 0
\end{equation}
Finally, we obtain the following result by using algebraic manipulations:
\begin{equation}
     \beta_u^{v*} = \sqrt{ \frac{\alpha_u^v(1+\zeta_u^vP_u)}{\xi^vB^v \log_2\left(1+ \frac{P_uG_u^v}{\sigma^2} \right)} }
\end{equation}
\end{IEEEproof} 

\begin{algorithm}[t!]
	\caption{\strut Lagrangian Relaxation-Based Communication Resource Allocation}
	\label{alg:profit}
	\begin{algorithmic}[1]
	
		\STATE{\textbf{Initialize:} $i=0$; $\beta_u^v(0)$, $\epsilon > 0$, $ \zeta_u^v (0), \xi^v(0)$, \text{and} $\varrho_l(0) >0, (l=1,2)$}, 
		
		\REPEAT
		\STATE{$i \leftarrow i +1$};
		\STATE{Update $\varrho_l(i+1),(l=1,2)$ according to (52)};
		\STATE{Update $\zeta_u^v (i+1)$, $\xi^v(i+1)$ according to (50), and (51)};
		\STATE{Update $\beta_u^v(i+1)$ according to (53)};
		\UNTIL{ $|\beta_u^v(i+1)-\beta_u^v(i)| \leq \epsilon$};
		
		\STATE{Then, set $\beta_u^v(t+1)$ as the desired solution}.
		
	\end{algorithmic}
	\label{Algorithm}
\end{algorithm}

\vspace{-0.5in} 
\subsection{ UAV-assisted MEC decision (UAD)}
For a given users tasks offloading decision $\boldsymbol{\alpha}$, and bandwidth allocation $\boldsymbol{\beta}$, the UAV-assisted MEC decision problem can be formulated as follows:

\begin{mini!}[2]                 
		{\boldsymbol{\theta, \gamma, \phi}}                               
		{\mathbf{Z}(\boldsymbol{\theta,\gamma, \phi})}{\label{opt:P1}}{\textbf{P3:}} 
		\addConstraint{ \sum\limits_{ u \in \mathcal{U}_v} \theta_u^{u \rightarrow v}  E_u^{v, \textrm{comp}} + \sum\limits_{ w \in \mathcal{V}, w \neq v}  E^{v \rightarrow w}}\nonumber
		\addConstraint{ + E^{v \rightarrow 0} + E^{v, \text{hov}} \leq E_v^{\mathbf{max}},  \forall v \in \mathcal{V},}
		\addConstraint{\sum\limits_{u \in \mathcal{U}_{v} } E_{u}^{v \rightarrow w, \textrm{comp}} + E^{w,\textrm{hov}} \leq E_{w}^{\mathbf{max}}, \forall w \in \mathcal{V},} \nonumber
		\addConstraint{ \ \ \ \ \ \ \ \ \ \ \ \ \ \ \ \ \ \ \ \ \ \ \ \ \ \ \ \ \   w \neq v,} 
		\addConstraint{\theta_u^{u \rightarrow v} + \sum_{\substack{w\in \mathcal{V}, \\ w\neq v}} \gamma_u^{v \rightarrow w} + \phi_u^{v \rightarrow 0} = 1, \forall u \in \mathcal{U}_v, } 
		\addConstraint{\theta_u^{ u\rightarrow v} \in \{0,1\}, \forall u \in \mathcal{U}_u, \forall v \in \mathcal{V},}
		\addConstraint{\gamma_u^{v \rightarrow w} \in \{0,1\}, \forall u \in \mathcal{U}_u, \forall v,w \in \mathcal{V},}
		\addConstraint{\phi_u^{v \rightarrow 0} \in \{0,1\}, \forall u \in \mathcal{U}_u, \forall v \in \mathcal{V},}
	\end{mini!}
From (57), we observe that decision vectors $\boldsymbol{\theta}$, $\boldsymbol{\gamma}$, and $\boldsymbol{\phi}$ are coupled in (57a) and (57d). Moreover, (57e), (57f), and (57g) are binary decision variables. Therefore, we conclude that \textbf{P3} is an NP-hard problem. Thus, in order to solve \textbf{P3}, we first relax the binary variables into continuous variables. Therefore, we can reformulate $\textbf{P3}$ in (57) as follows:       

\begin{mini!}[2]                 
		{\boldsymbol{\theta, \gamma, \phi}}                               
		{\mathbf{Z}(\boldsymbol{\theta,\gamma, \phi})}{\label{opt:UAD}}{\textbf{P31:}} 
		\addConstraint{ \sum\limits_{ u \in \mathcal{U}_v} \theta_u^{u \rightarrow v}  E_u^{v, \textrm{comp}} + \sum\limits_{ w \in \mathcal{V}, w \neq v}  E^{v \rightarrow w}}\nonumber
		\addConstraint{ + E^{v \rightarrow 0} + E^{v, \text{hov}} \leq E_v^{\mathbf{max}},  \forall v \in \mathcal{V},}
		\addConstraint{\sum\limits_{u \in \mathcal{U}_{v} } E_{u}^{v \rightarrow w, \textrm{comp}} + E^{w,\textrm{hov}} \leq E_{w}^{\mathbf{max}}, \forall w \in \mathcal{V},} \nonumber
		\addConstraint{ \ \ \ \ \ \ \ \ \ \ \ \ \ \ \ \ \ \ \ \ \ \ \ \ \ \ \ \ \   w \neq v,} 
		\addConstraint{\theta_u^{u \rightarrow v} + \sum_{\substack{w\in \mathcal{V}, \\ w\neq v}} \gamma_u^{v \rightarrow w} + \phi_u^{v \rightarrow 0} = 1, \forall u \in \mathcal{U}_v, } 
		\addConstraint{ 0 \leq \theta_u^{ u\rightarrow v} \leq 1, \forall u \in \mathcal{U}_u, \forall v \in \mathcal{V},}
		\addConstraint{ 0 \leq \gamma_u^{v \rightarrow w} \leq 1, \forall u \in \mathcal{U}_u, \forall v,w \in \mathcal{V},}
		\addConstraint{ 0 \leq \phi_u^{v \rightarrow 0} \leq 1, \forall u \in \mathcal{U}_u, \forall v \in \mathcal{V},}
	\end{mini!}

However, due to the coupling constraint in (58d), the problem state in (58) is hard to solve. Therefore, we apply the alternating direction method of multipliers (ADMM) approach \cite{boyd2011distributed} to solve the problem in a decentralized manner by decoupling constraint (58d). Let $\Omega_{\theta}$, $\Omega_{\gamma}$, and $\Omega_{\phi}$ be the feasible set of $\boldsymbol{\theta}$, $\boldsymbol{\gamma}$, and $ \boldsymbol{\phi}$, respectively.
The feasible set of $\boldsymbol{\theta}$ is define as follows:
\begin{equation}
    \label{feasible:theta}
    \begin{aligned}
        \Omega_{\theta} \triangleq \bigg\{ &\theta_v \in \mathbf{R}^{|\mathcal{U}_v|}\textbf{:} \sum_{u \in \mathcal{U}_v} \theta_u^{u \rightarrow v} E_u^{v,\textrm{comp}} + \sum\limits_{ w \in \mathcal{V}, w \neq v}  E^{v \rightarrow w}  \\
        &+ E_{v}^{\emph{hov}} \leq E_{v}^{\mathbf{max}},
        \sum\limits_{ u \in \mathcal {U}_v} \theta_u^{u \rightarrow v} f_u^v \leq  F_v^{\mathbf{max}}, 0 \leq  \theta_{u}^{u \rightarrow v} \leq  1 \bigg\},
    \end{aligned}
\end{equation}
Then, the feasible set for $\gamma_{w}$ is given by:
\begin{equation}
    \label{feasible:gamma_w}
    \begin{aligned}
        \Omega_{\gamma_{w}} \triangleq \bigg\{ &\gamma_{w} \in \mathbf{R}^{|\mathcal{U}_{v}|}\textbf{:} \sum\limits_{u \in U_{v} } E_{u}^{v \rightarrow w, \textrm{comp}} + E_{w}^{\textrm{hov}} \leq E_{w}^{\mathbf{max}},  \\
        &\sum\limits_{ u \in \mathcal {U}_v} \gamma_u^{v \rightarrow w} f_u^w \leq  F_w^{\mathbf{max}},0 \leq  \gamma_{u}^{v \rightarrow w} \leq  1 \bigg\}, \\
        &\forall w \in \mathcal{V}, w \neq v.
    \end{aligned}
\end{equation}
Here, $f_u^w$ is the computing capacity of UAV $w$ that is allocated to user $u$, and $F_w^{\mathbf{max}}$ is the maximum computing capacity of the server at UAV $w$. Then, $E_{w}^{\textrm{hov}}$ is the hovering energy of UAV $w$, which can be formulated as follows:

\begin{equation}
    E^{w,\textrm{hov}} = P^{w, \textrm{hov}} t^{w, \textrm{hov}}, \forall w \in \mathcal{V}, 
\end{equation}
where  
\begin{equation}
    t^{w,\textrm{hov}} = \max\limits_{u \in \mathcal{U}_v} \Bigg\{t_u^{v \rightarrow w,\textrm{comp}} \bigg\}.
\end{equation}
Finally, the feasible set for $\phi$ is given by:
\begin{equation}
    \label{feasible:phi}
    \begin{aligned}
        \Omega_{\phi} \triangleq \bigg\{ \phi \in \mathbf{R}^{|\mathcal{U}_{v}|}\textbf{:} \sum\limits_{ u \in \mathcal {U}_v} \phi_u^{v \rightarrow w} f_u^0 \leq  F_0^{\mathbf{max}},  0 \leq  \phi_{u}^{u \rightarrow v} \leq  1\bigg\},
    \end{aligned}
\end{equation}
where $f_u^0$ is the computing capacity of the server at the BS that  is allocated to user $u$, and $F_0^{\mathbf{max}}$ is the maximum computing capacity of the server at the BS.

Based on the feasible sets for $\boldsymbol{\theta, \gamma, \phi}$, we can re-write the problem \textbf{P31} as follows:
\begin{subequations}
\begin{align}
    \textbf{P32}: \underset{\boldsymbol{\theta, \gamma, \phi}}{ \min} & \; \; \mathbf{Z}( \boldsymbol{\theta, \gamma, \phi} )  \\
    \text{s.t.} & \nonumber\\
    & \boldsymbol{\theta} + \boldsymbol{\gamma} + \boldsymbol{\phi} = \boldsymbol{1},\\
    & \boldsymbol{\theta} \in \Omega_{\theta}, \boldsymbol{\gamma} \in \Omega_{\gamma}, \boldsymbol{\phi} \in \Omega_{\phi},
\end{align}
\end{subequations}
where, $\boldsymbol{1} $ is a $U_{v}$-vector with all elements equal to one.
Following ADMM method, we can drive the augmented Lagrangian function of the problem in (64) as follows:
\begin{equation}
\label{equa:lagrang-uade}
    \begin{aligned}
        \mathcal{L}_{\rho}(\boldsymbol{\theta, \gamma, \phi}) = &\mathbf{Z}( \boldsymbol{\theta, \gamma, \phi} )  + \boldsymbol{\lambda}^{T} ( \boldsymbol{\theta} + \boldsymbol{\gamma} + \boldsymbol{\phi} - \boldsymbol{1} ) \\
        &+ \frac{\rho}{2} \bigg|\bigg|\boldsymbol{\theta} + \boldsymbol{\gamma} + \boldsymbol{\phi} - \boldsymbol{1} \bigg|\bigg|_{2}^{2},  
    \end{aligned}
\end{equation}
where $|| \cdot || $ is the norm, $\boldsymbol{\lambda}$ is a Lagrangian multiplier of the constraint (64b), and $\rho$ is a constant. By sequentially updating variables with any order of $\boldsymbol{\theta, \gamma, \phi} $ and taking update of Lagrangian dual variable $\boldsymbol{\lambda}$ afterwards. The problem in (58) always guarantees a stationary solution due to the convexity of $\mathbf{Z}(\boldsymbol{\theta, \gamma, \phi})$, and linear constraints.
In this work, we assume that the UAV $v$ will make its decision first, then asks its neighboring UAVs and the central server at the BS afterward. Then, $\boldsymbol{\theta}$ at the UAV $v$ is updated as follows:
\begin{equation}
    \label{theta_update}
    \begin{aligned}
        \boldsymbol{\theta}^{(k+1)} &= \arg\min_{\boldsymbol{\theta}} \bigg\{  \mathbf{Z}( \boldsymbol{\theta} ) +  \boldsymbol{\lambda}^{(k),T} ( \boldsymbol{\theta} + \boldsymbol{\gamma}^{(k)} + \boldsymbol{\phi}^{(k)} - \boldsymbol{1} ) \\
        &+ \frac{\rho}{2} \bigg|\bigg|\boldsymbol{\theta} + \boldsymbol{\gamma}^{(k)} + \boldsymbol{\phi}^{(k)} - \boldsymbol{1} \bigg|\bigg|_{2}^{2}\bigg\}, \\
        & \forall \boldsymbol{\theta} \in \Omega_{\theta}.
    \end{aligned}
\end{equation}
After receiving the information of UAV $v$, the neighboring UAVs of $v$ will update the variable $\gamma_{w}$ as follows:
\begin{equation}
    \label{gamma_w_update}
    \begin{aligned}
        \boldsymbol{\gamma}_{w}^{(k+1)} &= \arg\min_{\boldsymbol{\gamma}_{w}} \bigg\{ \mathbf{Z}( \boldsymbol{\gamma} ) \\
        &+  \boldsymbol{\lambda}^{(k),T} \bigg( \boldsymbol{\theta}^{(k+1)} + \boldsymbol{\gamma}_{w} + \boldsymbol{\gamma}_{-w}^{(k)}  + \boldsymbol{\phi}^{(k)} - \boldsymbol{1} \bigg) \\
        &+ \frac{\rho}{2} \bigg|\bigg|\boldsymbol{\theta}^{(k+1)} + \boldsymbol{\gamma}_{w} +  \boldsymbol{\gamma}_{-w}^{(k)}  + \boldsymbol{\phi}^{(k)} - \boldsymbol{1} \bigg|\bigg|_{2}^{2}\bigg\},  
    \end{aligned}
\end{equation}
Then, the terrestrial BS updates $\phi$ as follows:
\begin{equation}
    \label{phi_update}
    \begin{aligned}
        \boldsymbol{\phi}^{(k+1)} &= \arg\min_{\boldsymbol{\phi}} \bigg\{ \mathbf{Z}( \boldsymbol{\phi} ) +  \boldsymbol{\lambda}^{(k),T} ( \boldsymbol{\theta}^{(k+1)} + \boldsymbol{\gamma}^{(k+1)} + \boldsymbol{\phi} - \boldsymbol{1} ) \\
        &+ \frac{\rho}{2} \bigg|\bigg|\boldsymbol{\theta}^{(k+1)} + \boldsymbol{\gamma}^{(k+1)} + \boldsymbol{\phi} - \boldsymbol{1} \bigg|\bigg|_{2}^{2}\bigg\},  
    \end{aligned}
\end{equation}
 Finally, by using the updated $ \boldsymbol{\theta}^{(k+1)}$, $ \boldsymbol{\gamma}_{w}^{(k+1)}$, and $ \boldsymbol{\phi}^{(k+1)}$, the dual variable $\boldsymbol{\lambda}$ can be updated as follows:
\begin{equation}
    \label{lambda_update}
    \begin{aligned}
        \boldsymbol{\lambda}^{k+1} = & \boldsymbol{\lambda}^{k+1} + \rho \bigg( \boldsymbol{\theta}^{(k+1)} + \boldsymbol{\gamma}^{(k+1)} + \boldsymbol{\phi}^{(k+1)} - \boldsymbol{1} \bigg).
    \end{aligned}
\end{equation}
A summary of the proposed ADMM-based UAV-assisted MEC decision is presented in Algorithm 2.

\begin{algorithm}[t]
	\caption{ ADMM-Based UAV-assisted MEC Decision}
	\label{alg:admm_based}
	\begin{algorithmic}[1]
	 \renewcommand{\algorithmicrequire}{\textbf{Input:}}
        \renewcommand{\algorithmicensure}{\textbf{Output:}}
	    \REQUIRE $ \mathcal{V}, \ \mathcal{U}$;
	    \ENSURE $\boldsymbol{\theta, \gamma, \phi }$;
		\STATE{\textbf{Initialize:} $t \leftarrow 0$; $ \boldsymbol{\theta}^{(0)} \leftarrow 0 $, $ \boldsymbol{\gamma^{0}} \leftarrow 0$,
		$ \boldsymbol{\phi^{(0)}} \leftarrow 0$,
		$\boldsymbol{\lambda^{(0)}} \leftarrow 0,$
		\text{and} $\rho = 10.0$};
		
		\REPEAT
		\STATE{ The UAV $v$  updates $\boldsymbol{\theta}^{(t+1)}, \forall u\in U_{v}$ according to \eqref{theta_update}: 
		\begin{equation*}
        \begin{aligned}
        \boldsymbol{\theta}^{(k+1)} &= \arg\min_{\boldsymbol{\theta}} \bigg\{  \mathbf{Z}( \boldsymbol{\theta} ) +  \boldsymbol{\lambda}^{(k),T} ( \boldsymbol{\theta} + \boldsymbol{\gamma}^{(k)} + \boldsymbol{\phi}^{(k)} - \boldsymbol{1} ) \\
        &+ \frac{\rho}{2} \bigg|\bigg|\boldsymbol{\theta} + \boldsymbol{\gamma}^{(k)} + \boldsymbol{\phi}^{(k)} - \boldsymbol{1} \bigg|\bigg|_{2}^{2}\bigg\}, \\
        & \forall \boldsymbol{\theta} \in \Omega_{\theta};
       \end{aligned}
       \end{equation*}}
		\STATE{Each neighboring UAV $w \in \mathcal{V}, w \neq v$ updates its decision $\boldsymbol{\gamma}_{w \in \mathcal{V}} $ parallelly  according to \eqref{gamma_w_update}:
		\begin{equation*}
         \begin{aligned}
        \boldsymbol{\gamma}_{w}^{(k+1)} &= \arg\min_{\boldsymbol{\gamma}_{w}} \bigg\{ \mathbf{Z}( \boldsymbol{\gamma} ) \\
        &+  \boldsymbol{\lambda}^{(k),T} \bigg( \boldsymbol{\theta}^{(k+1)} + \boldsymbol{\gamma}_{w} + \boldsymbol{\gamma}_{-w}^{(k)}  + \boldsymbol{\phi}^{(k)} - \boldsymbol{1} \bigg) \\
        &+ \frac{\rho}{2} \bigg|\bigg|\boldsymbol{\theta}^{(k+1)} + \boldsymbol{\gamma}_{w} +  \boldsymbol{\gamma}_{-w}^{(k)}  + \boldsymbol{\phi}^{(k)} - \boldsymbol{1} \bigg|\bigg|_{2}^{2}\bigg\}; 
        \end{aligned}
        \end{equation*}}
		\STATE{The terrestrial BS updates the value of its decision $\boldsymbol{\phi}^{(t+1)}$ according to \eqref{phi_update}:
		\begin{equation*}
        \begin{aligned}
        \boldsymbol{\phi}^{(k+1)} &= \arg\min_{\boldsymbol{\phi}} \bigg\{ \mathbf{Z}( \boldsymbol{\phi} ) +  \boldsymbol{\lambda}^{(k),T} ( \boldsymbol{\theta}^{(k+1)} + \boldsymbol{\gamma}^{(k+1)} + \boldsymbol{\phi} \\
        &  - \boldsymbol{1})+ \frac{\rho}{2} \bigg|\bigg|\boldsymbol{\theta}^{(k+1)} + \boldsymbol{\gamma}^{(k+1)} + \boldsymbol{\phi} - \boldsymbol{1} \bigg|\bigg|_{2}^{2}\bigg\};  
       \end{aligned}
       \end{equation*}}
		\STATE{ The terrestrial BS and Neighbor UAVs feed back the information about it decision to UAV $v$};
		\STATE{ After receiving the information about the neighboring UAVs and MBS, the UAV $v$ then updates the Lagrangian dual variable  according to \eqref{lambda_update}:
		\begin{equation*}
         \begin{aligned}
        \boldsymbol{\lambda}^{k+1} = & \boldsymbol{\lambda}^{k+1} + \rho \bigg( \boldsymbol{\theta}^{(k+1)} + \boldsymbol{\gamma}^{(k+1)} + \boldsymbol{\phi}^{(k+1)} - \boldsymbol{1} \bigg);
        \end{aligned}
        \end{equation*}}
		\STATE{$t \leftarrow t +1$};
		\UNTIL{ $|| \boldsymbol{(\theta}^{(t+1)} + \boldsymbol{\gamma}^{(t+1)} + \boldsymbol{\phi}^{(t+1)}) - (\boldsymbol{\theta}^{(t)} + \boldsymbol{\gamma}^{(t)}+\boldsymbol{\phi}^{(t)} ) || \leq \epsilon_{\emph{pri}} \cap || \boldsymbol{\lambda}^{(t+1)} - \boldsymbol{\lambda}^{(t)} || \leq \epsilon_{\emph{dual}}$  };
		\STATE{Then, set $\boldsymbol{\theta, \gamma, \phi} $ as the desired solution}.
	\end{algorithmic}
	\label{Algorithm}
\end{algorithm}

\section{Simulation Results}
\label{simu}
In this section, we evaluate the performance of our proposed algorithm in the considered collaborative multi-UAV-assisted MEC system that integrates with MEC-enabled terrestrial BS. To evaluate our proposed algorithms, we use PYTHON programming language, and SCS solver in the CVXPY toolkit.

\subsection{Simulation Setup}
We randomly deploy 10 UAVs and $[5, 50]$ mobile users within a $400$ m  $\times$ $400$ m rectangular region. We assume that UAVs are hovering at a fixed altitude of $50$ m. The BS's location has been set at $(0,0,0)$. In addition, the Rician channel fading and the free-space path loss model are adopted. Finally, the main parameters employed in our simulation are listed in table \ref{tab:table_simulation}.

\begin{table}[t!]
	\caption{Summary of Notations.}
	\textbf{\label{tab:table_simulation}} 
	\renewcommand\arraystretch{1}
	\begin{center}
		\begin{tabular}{|m{1.5cm}|m{2cm}||m{1.5cm}|m{2cm}|}
			\hline
			\hfil \textbf{Parameter} & \hfil \textbf{Value} & \hfil \textbf{Parameter} & \hfil \textbf{Value} \\ \hline \hline
			\ \hfil $G_0$ & -50 dB  & \hfil $P_u$ &  23 dBm \\ \hline
			\ \hfil $\sigma^2$ & -174 dBm  & \hfil $S_u$ &  [100, 500] MB  \\ \hline
            \ \hfil $C_u$ & [10, 50] Cycles  & \hfil $f_u$ &  [0.5, 3] MHz \\ \hline
			\ \hfil $\kappa$ & 5 $\times$ $10^{-27}$  & \hfil $B^v$ &  3 MHz  \\ \hline \ \hfil $F_v^{\mathbf{max}}$ & [1, 3.5] MHz  & \hfil $\varpi$ &  2  \\ \hline
			\ \hfil $P_v$ & 30 dBm & \hfil $B^{v \rightarrow w}$ &  1.7 MHz  \\ \hline
		    \ \hfil $\eta$ & 30 N \cite{monwar2018optimized} & \hfil $\varphi_v$ & 70 \% \cite{monwar2018optimized} \\ \hline
		   \ \hfil $j$ & 4 \cite{monwar2018optimized}  & \hfil $r$ &  0.254 \cite{monwar2018optimized}  \\ \hline
		   \ \hfil $\varkappa$ & 1.225 kg/m$^3$ & \hfil $E_u^{\mathbf{max}}$ &  100 kJ  \\ \hline
		   \ \hfil $E_v^{\mathbf{max}}$ & 500 kJ & \hfil $F_0^{\mathbf{max}}$ & [2, 4] MHz \\ \hline
		    \ \hfil $B_c^{\textrm{mm}}$ & 28 GHz & \hfil $B_{\textrm{mm}}^{v \rightarrow 0}$ &  1.8 MHz  \\ \hline
		    \ \hfil $P^{v \rightarrow 0}$ & 30 dBm & \hfil $\epsilon_{pri}$ & $10^{-4}$  \\ \hline
		    \ \hfil $\epsilon_{dual}$ & $10^{-5}$ & \hfil $\rho$ & [1, 5, 10, 15]  \\ \hline
		\end{tabular}
		\label{tab1}
	\end{center}
\end{table}

\subsection{Results}
We now focus on the performance assessment of our proposed algorithm. We compare the performance of our proposed approach to the performance of the following existing schemes:

\begin{itemize}
  \item \emph{Centralized algorithm}: In this baseline, the BS serves as a central coordinator that needs complete information (i.e., information about task profiles, UAVs' computing capacity,  channel state information (SCI) between UAVs and users) as inputs to optimize the system utility. It should be noted that the centralized algorithm is intended to solve the relaxed problem described in (58).    
  \item \emph{Greedy algorithm}: Based on the channel state information, the UAV $v$ needs to sort the neighboring UAV indices by decreasing order with respect to the available bandwidth between it and neighboring UAV $w \in \mathcal{V}, w \neq v$. Then, UAV $v$ takes into account the available computing resources of the neighboring UAVs and the demand (i.e., the required CPU cycles to complete the task) of computing tasks of its associated users. If the available computing resource at the UAV $w$ is satisfied with the demand of the computing task of user $u$, UAV $v$ will forward the user's task to the neighboring UAV $w$. This process is repeated until all of the users' computing tasks at UAV $v$ have been forwarded or the computing resources of neighboring UAVs have been consumed. If there is still computing tasks that cannot be forwarded to the neighboring UAVs, UAV $v$ will forward them to the terrestrial BS.

  \item \emph{Exhaustive search}: The original problem in (57) is a  binary decision problem with multi-dimensional variables. The combination of the variables is a binary matrix, with each element being either one or zero.  Furthermore, the binary matrix satisfies the following two facts: i)  summation of the column vector is less than or equal to one as stated in (57d), and ii) summation of the row vector is upper bounded by maximum capacity (57b), (57c). In this exhaustive search approach, firstly all of the possible outcomes of the matrix are generated. Then, amongst the elements of the outcome set, we try to find the smallest one, which is the best solution for the original problem (the non-relaxed problem described in (57)).    
  
\end{itemize}

\begin{figure}[t!]
    \centering
    \includegraphics[width=\linewidth]{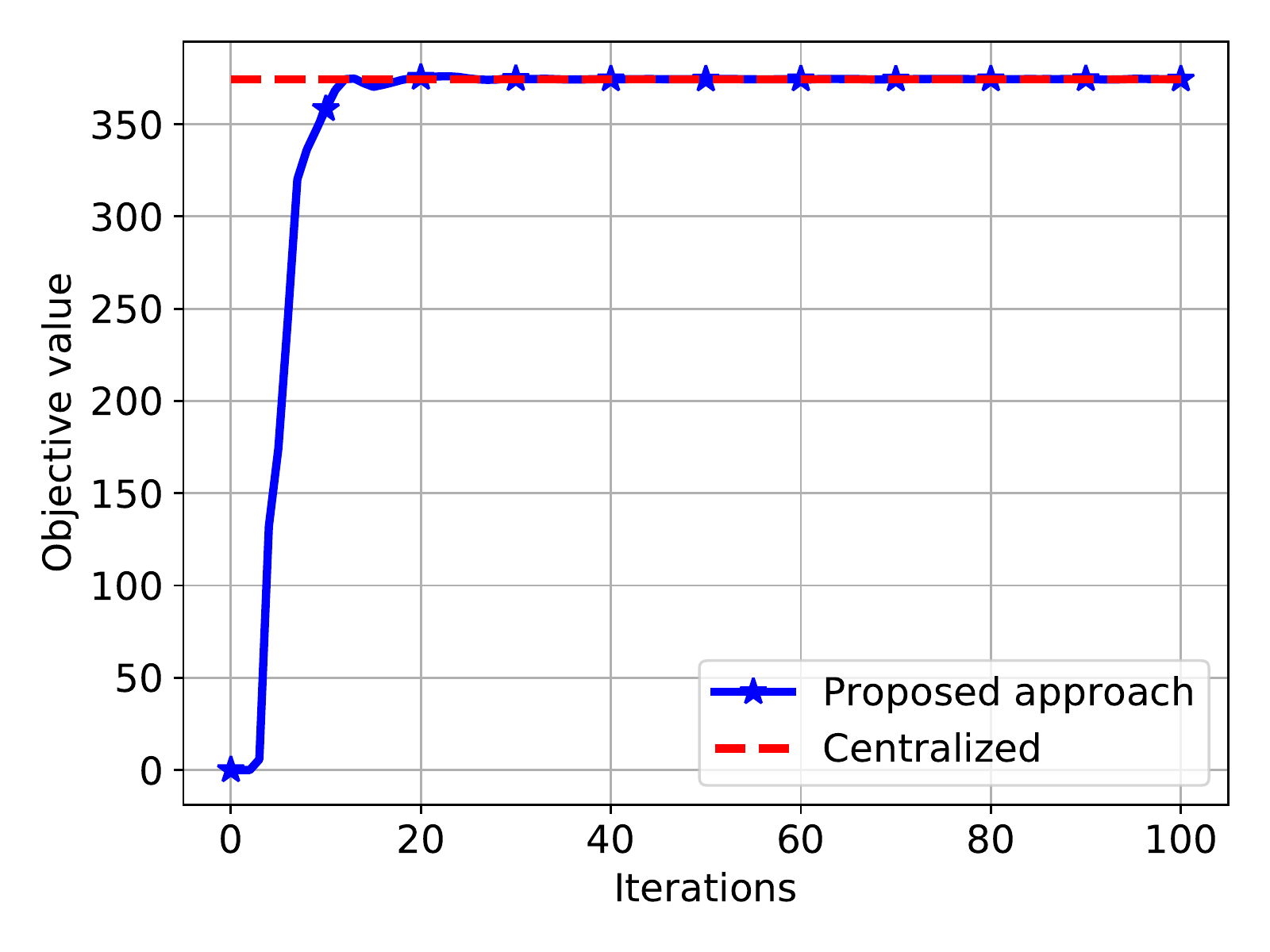}
    \caption{Convergence of the proposed algorithm.}
    \label{fig:admm}
\end{figure}

\begin{figure}[t!] 
    \centering
    \includegraphics[width=\linewidth]{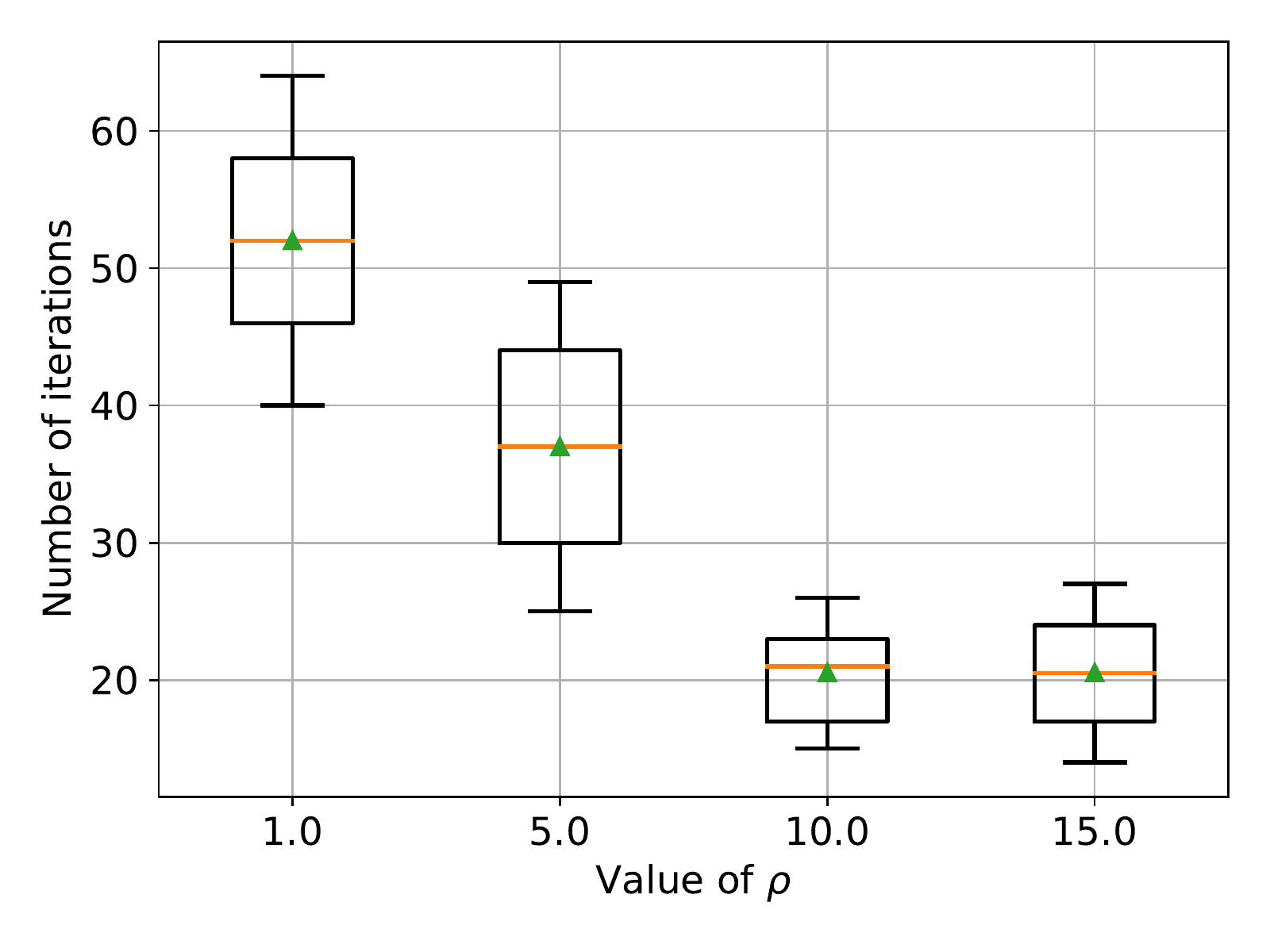}
    \caption{Number of iterations versus the value of $\rho$.}
    \label{fig2}
\end{figure}

The convergence of the proposed algorithm is presented in Fig.~\ref{fig:admm}. When the number of iterations is smaller than 20, our objective function (i.e., total latency of users) violates the resource limitation constraints, as shown in Fig.~\ref{fig:admm}. From the figure, we can observe that after 20 iterations, our proposed algorithm converges to the specific value, also known as stationary point, which is regarded as the optimal solution that does not violate the resource limitation constraints. In addition, we observe that the proposed algorithm converges to the optimal solution within 20 iterations. Thus, our algorithm is practical. Then, Fig.~\ref{fig2} demonstrates the influence of the penalty value $\rho$ on the number of iterations required to reach to the optimal solution. As can be seen in the figure, increasing the value of the penalty parameter $\rho$ reduces the number of iterations required to reach to the optimal solution. The parameter $\rho$ determines how the augmented Lagrangian term is treated inside $||.||_2^2$. Moreover, it tightens the equality constraint, and it is related to the convergence rate of ADMM \cite{boyd2004convex}. Depending on the problem, we must carefully choose the value of the parameter $\rho$. The most typical value of $\rho$ is 0.5 or 1.0. However, in this paper, we have an equality constraint that states that the demand of the user must be severed (58d). Thus, we tune the parameter to find the best value to deal with the convergence rate of the algorithm as well as the complementary slackness of the dual constraint $\lambda$.    

\begin{figure}[t!]
	\centering
	\includegraphics[width=\linewidth]{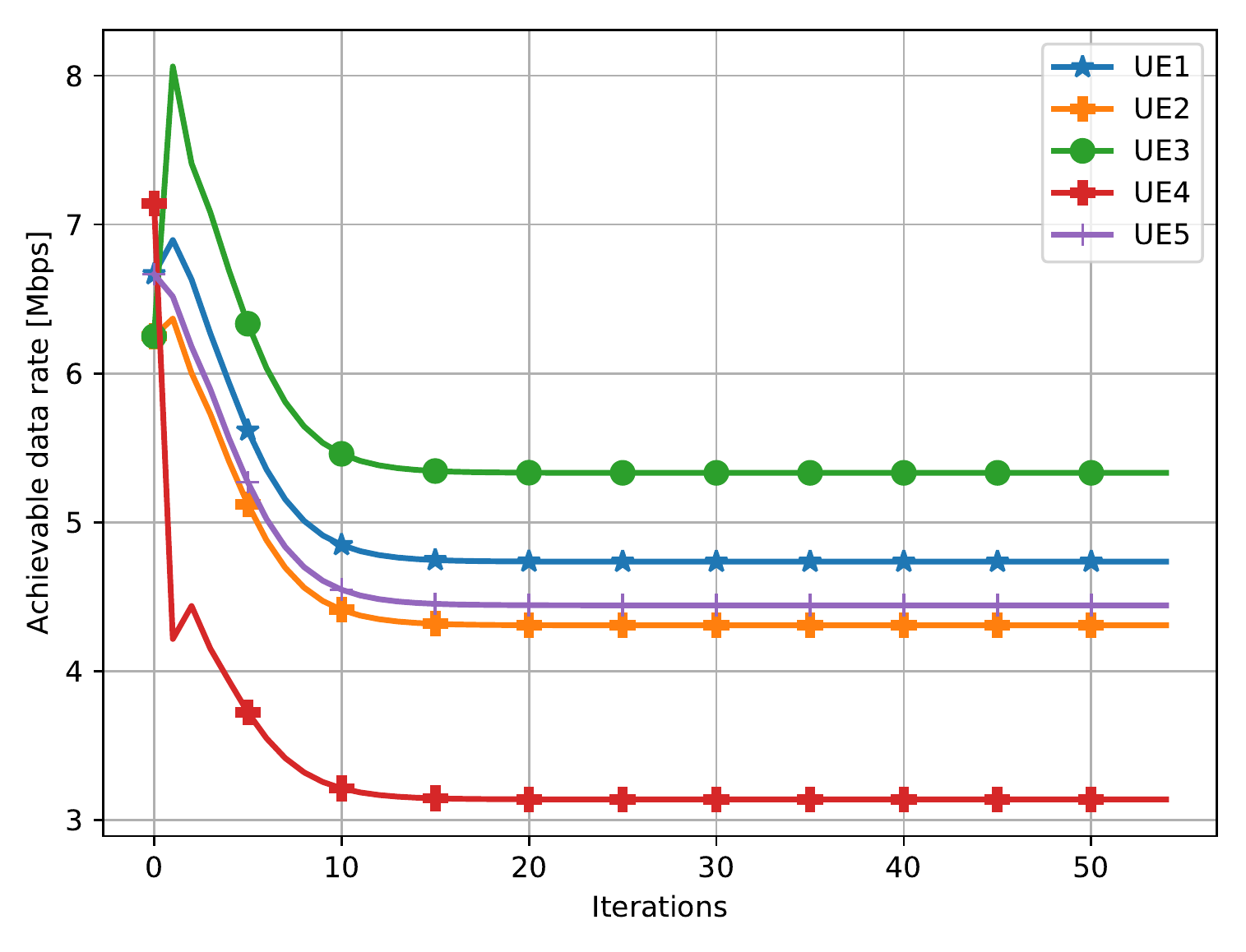}
	\caption{The convergence performance of communication resource allocation.}
	\label{fig:convergenc_of_dual_decompose}
\end{figure}

In Fig.~\ref{fig:convergenc_of_dual_decompose}, we show the achievable data rate of each user when the total number of users in the network is 5. From this figure, we observe that the achievable data rate of each user converges to the optimal solution within a few iterations. This shows the effectiveness of our proposed Lagrangian relaxation-based algorithm. Furthermore, compared to other users in the network, we can see that UE $3$ has the highest achievable data rate. This is because the achievable channel gain between UAV and UE $3$ is the highest, and UE $3$ receives the largest share of the system bandwidth since its offloads a larger portion of its computing task to the associated UAV.

\begin{figure}[h]
	\centering
	\includegraphics[width=\linewidth]{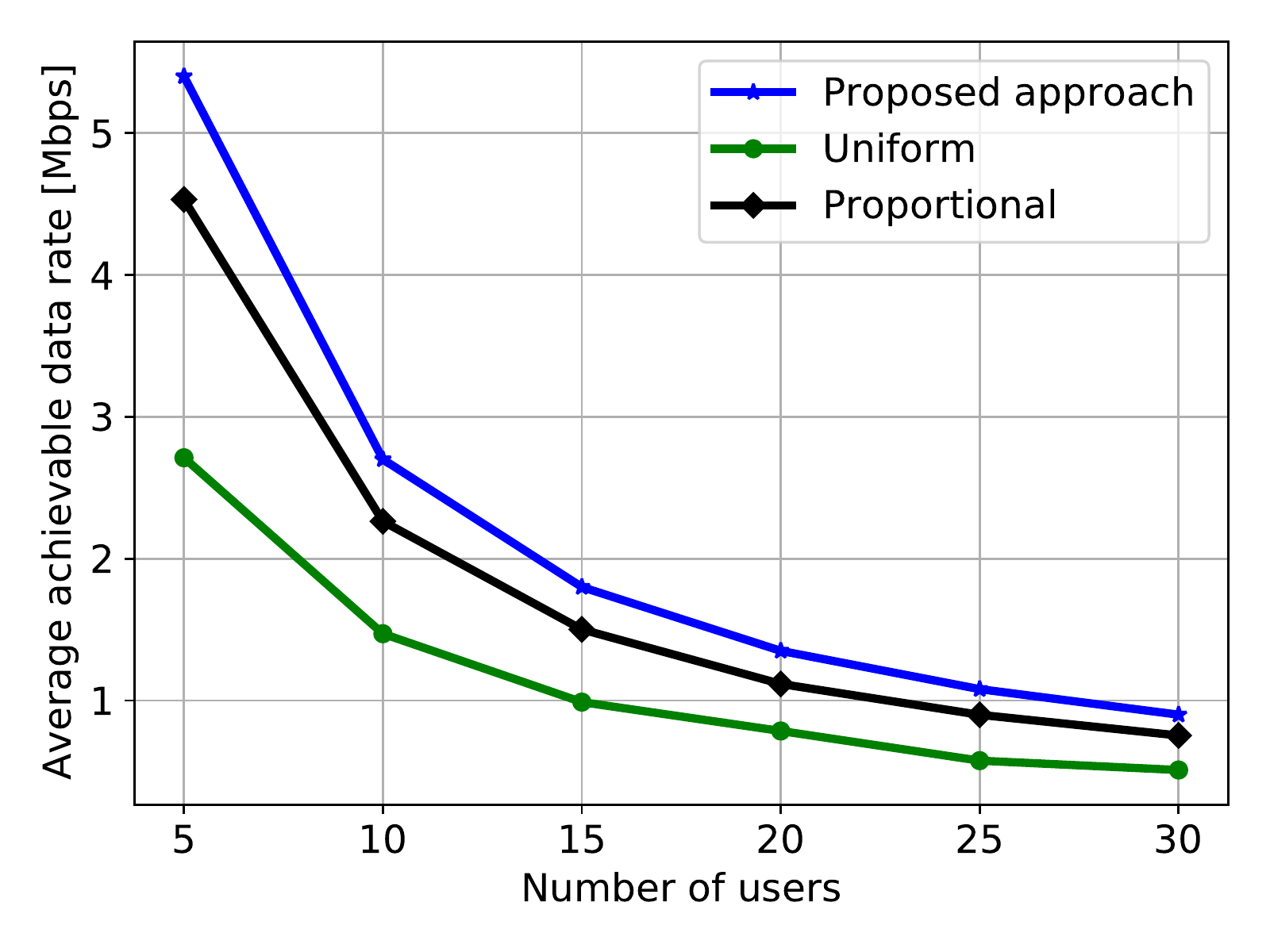}
	\caption{Average achievable data rate under different number of users.}
	\label{fig:avg_data_rate}
\end{figure}

\begin{figure}[t!]
    \centering
    \includegraphics[width=\linewidth]{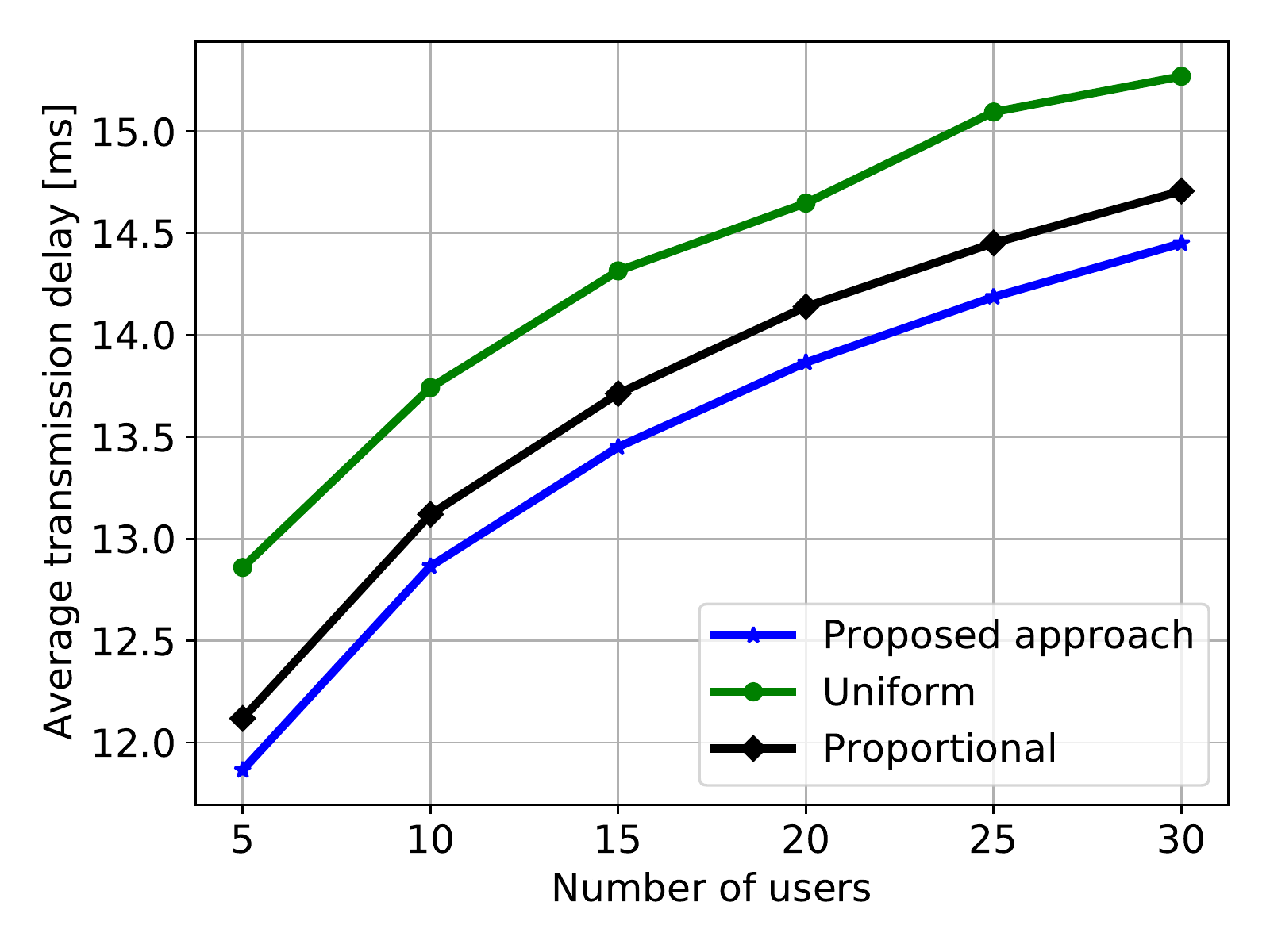}
    \caption{Average transmission latency under different number of users.}
    \label{fig: latency}
\end{figure}

Fig~\ref{fig:avg_data_rate} shows the average achievable data rate for different network sizes. Moreover, we also compare the average achievable data rate under our proposed algorithm, i.e., the Lagrangian relaxation method, with the uniform and proportional resource allocation schemes. Under a uniform resource allocation scheme, the available bandwidth at each UAV is uniformly allocated to its associated users. Then, in the proportional allocation scheme, each UAV allocates its bandwidth to each user based on the size of the offloaded data, as follows:
\begin{equation}
    \beta_u^v = \frac{\alpha_u^v}{ \sum\limits_{q \in  \mathcal{U}_v}  \alpha_q^v} B^v, \forall u \in \mathcal{U}_v, \forall v \in \mathcal{V}. 
\end{equation}
From Fig.~\ref{fig:avg_data_rate}, we observe that the average achievable data rate resulting from our proposed algorithm is higher than that of the uniform and proportional scheme. In addition, as seen in Fig.~\ref{fig:avg_data_rate}, the average achievable data rate decreases as the number of users in the network grows. This is because each UAV has limited bandwidth for the associated users. Moreover, in Fig.~\ref{fig: latency}, we show the average transmission latency under different network sizes. Fig.~\ref{fig: latency} shows that as the number of users in the network grows, the average transmission latency/delay increases. The reason is that in (12), we can clearly see that the transmission latency is inversely proportional to the achievable data rate. In other words, the latency decreases when the transmission rate increases. From Fig.~\ref{fig:avg_data_rate}, we already observed that the transmission rate is dramatically decreasing with the increasing number of users in the network. Therefore, the latency will decrease with the increasing number of users as, as we have seen in Fig.~\ref{fig: latency}. Moreover, Fig.~\ref{fig: latency} demonstrates that the average transmission latency under our proposed algorithm is the lowest when compared to uniform and proportional allocation schemes. Therefore, we can conclude that our proposed approach outperforms other schemes.
  
\begin{figure}[h]
	\centering
	\includegraphics[width=\linewidth]{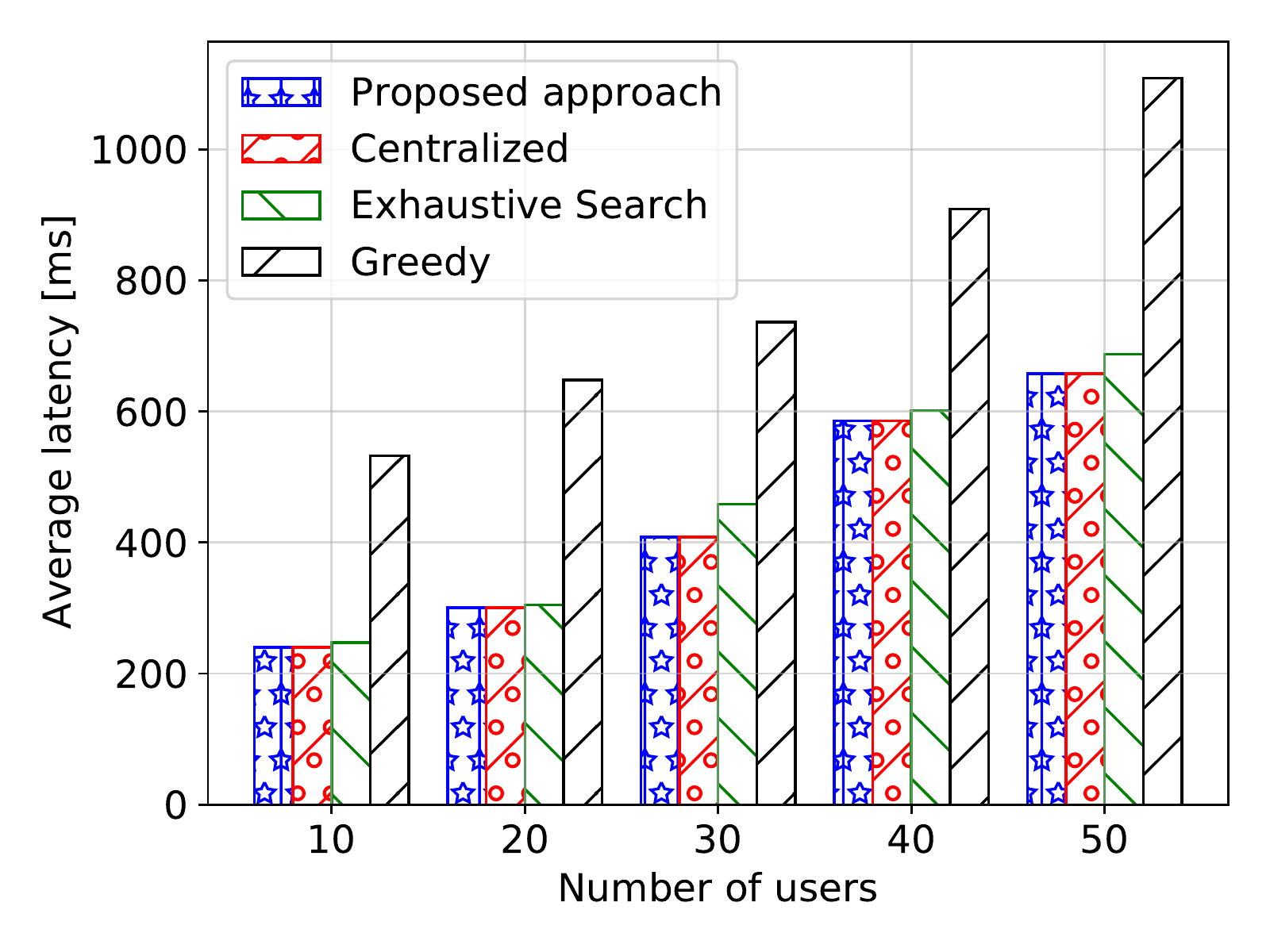}
	\caption{Comparison of average total latency.}
	\label{compare}
\end{figure}

Fig.~\ref{compare} shows the average total network latency (i.e., objective value) experienced by the users. Moreover, this figure compares the performance of our proposed algorithm with the centralized, greedy, and exhaustive search schemes. As an example, when there are 50 users in the network, the average latency is 657.394 ms (proposed approach), 657.394 ms (centralized), 687.094 ms (exhaustive search), and 1108.806 ms (greedy), respectively. From the results above, we observe that the average latency under our proposed algorithm is 40.7\% and 4.3\% less than that of the greedy and exhaustive search schemes. In the greedy scheme, we only consider the factor of available bandwidth between UAVs while deciding whether or not the computing tasks of its associated users are forwarded to the neighboring UAVs, and we omitted the full utilization of computation resource at the neighboring UAVs. Therefore, the greedy scheme performs worse than our proposed approach. On the other hand, the goal of the exhaustive search method is to concentrate on offloading decision-making without taking into account computing resource (i.e., capacity) allocation. Furthermore, the offloading decisions are binary decision variables, i.e., the computing tasks of the users are offloaded to the single neighboring UAV, which is not like offloaded to the multiple locations (i.e., neighboring UAVs and BS) as in our relaxed problem. Thus, the computing capacity of some neighboring UAVs might not be fully utilized. In contrast to the techniques stated above, our proposed approach is more flexible because the computing tasks are separable and offloaded to multiple neighboring UAVs. Thus, a UAV's utilization, i.e., the utilization of all neighboring UAVs' computing capacity, may simply be achieved by offloading a fraction of its users' tasks to all UAVs. Furthermore, the average latency under our proposed algorithm is the same as the centralized solution (i.e., optimal solution). Finally, as seen in Fig.~\ref{compare}, average latency increases as the number of users in the network increases.

\begin{figure}[t!]
    \centering
    \includegraphics[width=\linewidth]{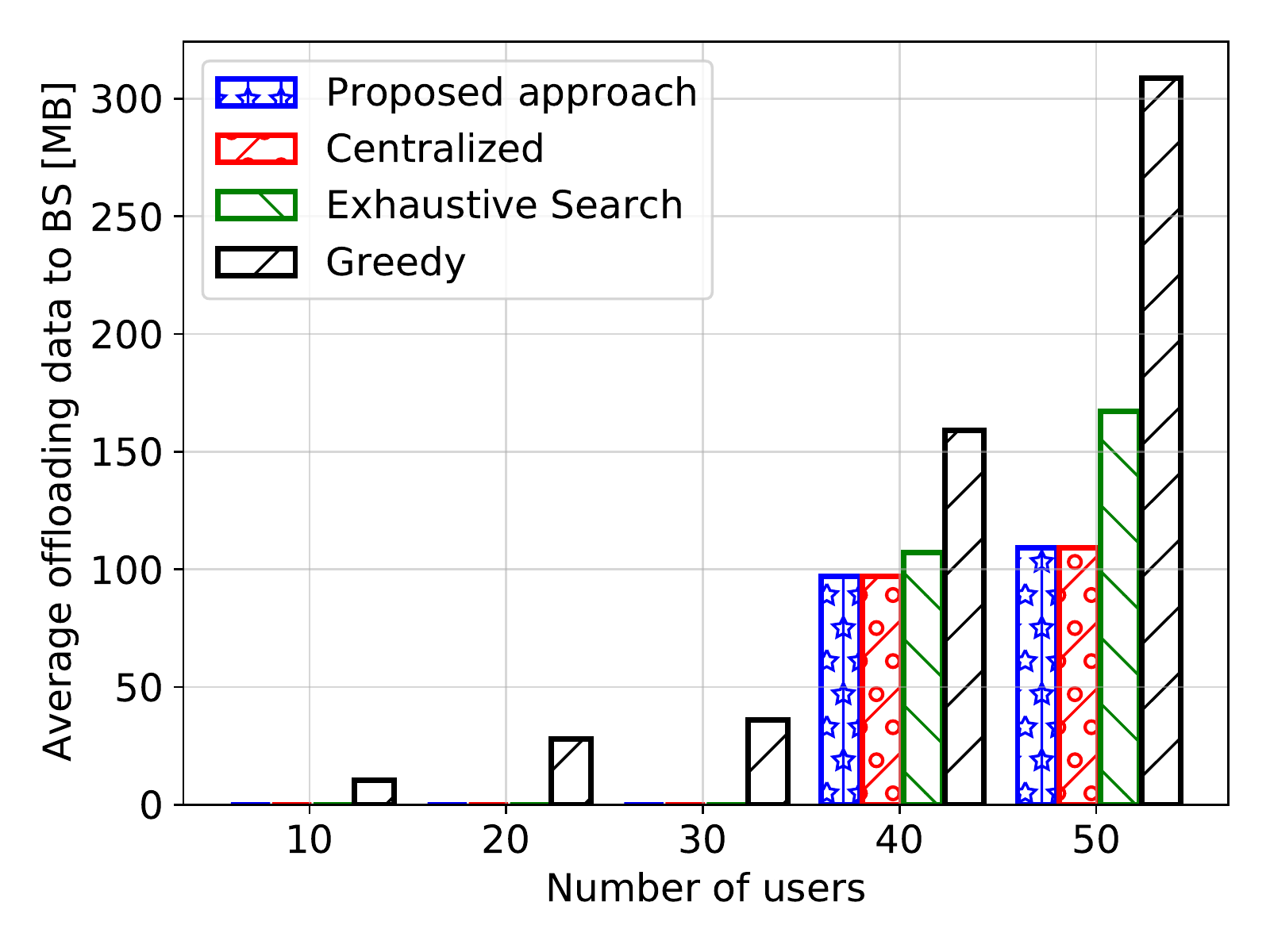}
    \caption{Comparison of Average Offloading Data to the terrestrial BS.}
    \label{fig:my_label}
\end{figure}

\begin{figure}[t!]
	\centering
	\includegraphics[width=\linewidth]{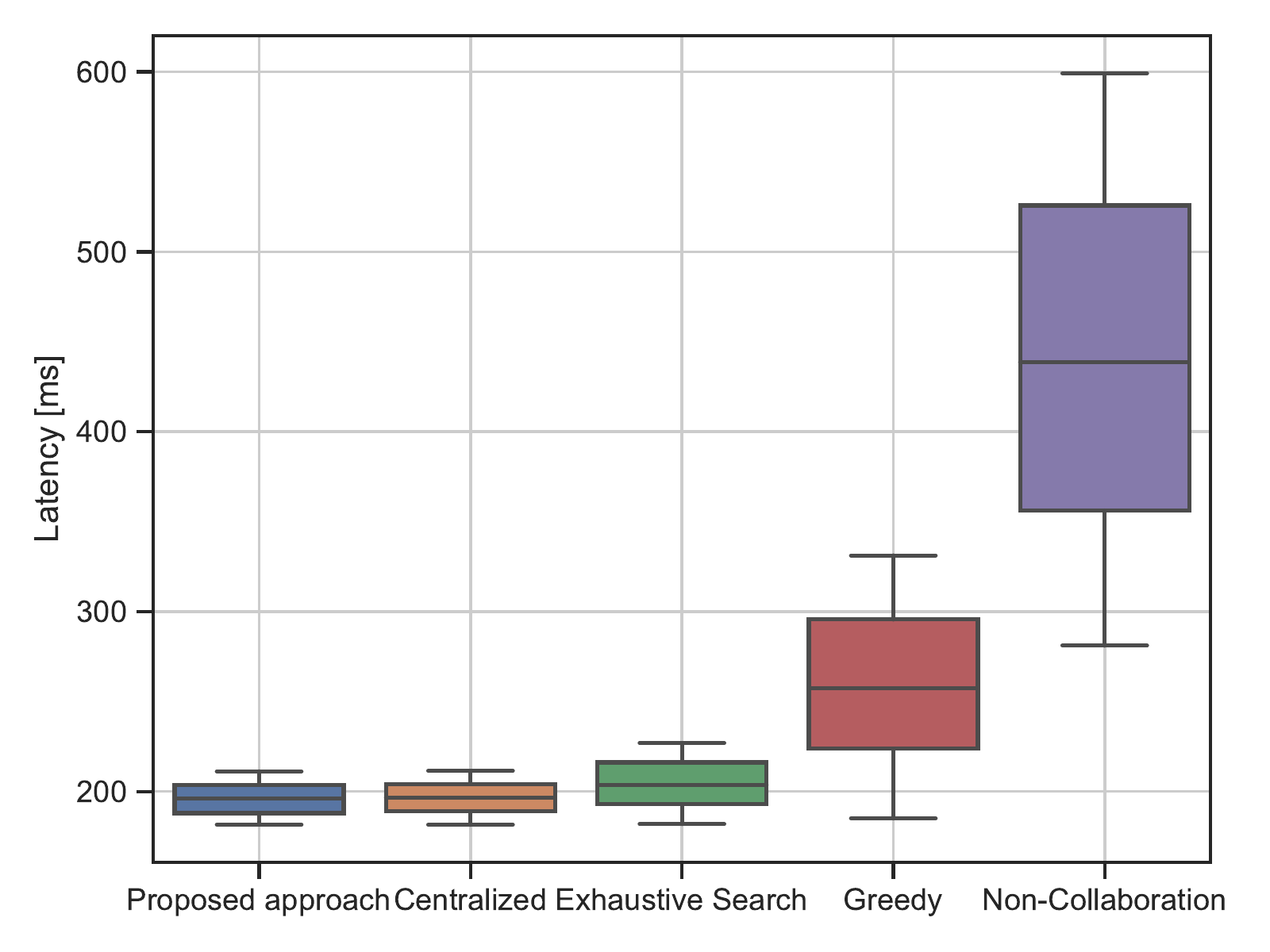}
	\caption{Performance comparison with fixed number of $UAV = 3$, number of UEs 10. }
	\label{fig:with}
\end{figure}

Fig.~\ref{fig:my_label} shows the average amount of data offloaded to the terrestrial BS via the greedy algorithm, exhaustive search, centralized scheme, and proposed algorithm (proposed approach), respectively. In this figure, when the number of users is not more than 30, we can observe that the average data offloaded to the BS under our proposed algorithm, centralized, and exhaustive search schemes are negligible. This is because the computing tasks offloaded by the users can be executed within the deadline on the associated UAV and neighboring UAVs without relying on the terrestrial BS. Further, it can be seen that as the number of users grows, the average amount of data offloaded to BS also increases due to the limited ability of UAVs to execute the offloaded tasks from their users. Therefore, a larger portion of the computing tasks from the users will be offloaded to the BS.

Finally, Fig.~\ref{fig:with} depicts the users' latency at a fixed network size with 3 UAVs and 10 users. In addition, we compare the performance of our proposed approach to that of other existing schemes. From the figure, we can observe that the users' latency under our proposed approach is comparable to that of the centralized algorithm. Furthermore, the users' latency is 220.234 ms (proposed approach), 247.2712 ms (exhausted search), 345.7263 ms (greedy), and 599.6213 ms (non-collaboration), respectively. From the result, we notice that the users' latency is the highest under the non-collaboration scheme. In the non-collaboration scheme, when a UAV's computing capacity is insufficient to execute tasks of the associated users, the UAV forwarded the offloaded tasks of users straight to the BS. As the BS is far away from the UAV, the UAV-to-BS transmission latency will be higher when compared to the UAV-to-UAV and user-to-UAV transmission latency. Therefore, users' latency under non-collaboration is higher than that of the other schemes.

\section{Conclusion}
\label{con} 

In this paper, we have proposed a collaborative multi-UAV-assisted MEC system that integrates with the MEC-enabled terrestrial base station. Then, we have studied the latency minimization problem by optimizing the offloading decision, communication, and computing resource allocation under the energy budget of both users and UAVs. Next, we have shown that the formulated problem is mixed-integer, nonlinear, and non-convex. Therefore, to be tractable, we have decomposed the formulated problem into three subproblems: users tasks offloading decision problem, communication resource allocation problem, and UAV-assisted MEC decision problem. Then, we solved users tasks offloading decision problem by using the SCS solver in CVXPY. Moreover, we have deployed Lagrangian relaxation and ADMM approaches to solve the communication resource allocation and UAV-assisted MEC decision problem. Finally, we have provided the extensive numerical results to show the effectiveness of our proposed algorithms. From the results, it is clear that our proposal reduced the latency experienced by the users and achieves fast convergence. An important future work future work would be to consider the integration of our system with a satellite network.

\bibliographystyle{IEEEtran}
\bibliography{Colla}

%




\end{document}